\documentclass{aa}

\usepackage{graphicx}
\usepackage{txfonts}
\usepackage{tabularx}  
\usepackage{array} 
\usepackage{multirow} 

\begin{document} 

   \title{Hidden metallic iron in amorphous silicate dust?}
   \subtitle{Insights from condensation experiments and mid-infrared spectroscopy}

   \author{H. Enomoto
          \inst{1}
          \and
          A. Takigawa
          \inst{1}
          \and
          H. Chihara
          \inst{2}
          \and
          C. Koike\inst{2}
          }

   \institute{Department of Earth and Planetary Science, The University of Tokyo, 7-3-1 Hongo, Tokyo 113-0033, Japan\\
              \email{hana@eps.s.u-tokyo.ac.jp}
         \and
             Department of Environmental Science and Technology Science, Osaka Sangyo University, Osaka 574-8530, Japan\\
             }

   \date{Received}
 
  \abstract
   {Amorphous silicate dust is a major component in the interstellar and circumstellar dust formed in the outflow of asymptotic giant branch (AGB) stars. Although iron depletion is observed in the interstellar medium, the exact form and fraction of iron in solids remains a matter of debate. In particular, it is unclear whether the amorphous silicate dust around AGB stars contains metallic iron.}
   {We aimed to provide optical constants of amorphous silicate nanoparticles and examine the effects of metallic iron on their spectral features to better constrain the dust properties. We did this by producing amorphous silicate nanoparticles with and without metallic cores.}
   {We performed condensation experiments using an induction thermal plasma system to produce dust analogs of the CI chondritic composition in Mg–Ca–Na–Al–Si–Fe–Ni–O and Mg–Ca–Na–Al–Si–O systems. We measured the absorbance and reflectance of the samples, observed the structure of the products, and determined the optical constants.}
   {Two types of amorphous silicate nanoparticles ($\phi 10-200 \mathrm{\, nm}$) with nearly CI chondritic compositions were produced: one contained kamacite (Fe$_{0.9}$Ni$_{0.1}$) cores with a diameter ratio ranging from 0–0.87 (average~0.50), and the other was iron-free homogeneous amorphous silicate. The amorphous silicates of the CI chondritic composition with various-sized metallic cores may be prevalent in circumstellar and interstellar dust.}
   {}

\titlerunning {Hidden metallic iron in amorphous silicate dust? Insights from condensation experiments and MIR spectroscopy}

   \maketitle

\section{Introduction}
  Infrared (IR) spectroscopic observations have revealed that amorphous silicate dust is a dominant refractory dust in circumstellar and interstellar environments. At least 97.8\% of silicate dust (by mass) is amorphous in the diffuse interstellar medium (ISM; \citealt{Kemper2004, Kemper2005}). Amorphous silicate ubiquitously exists in protoplanetary disks and interplanetary dust particles as well \citep{Henning2010}.
  
  Clear evidence of amorphous silicate formation was observed in outflows of asymptotic giant branch (AGB) stars, which are evolved stars with low to intermediate masses ($0.8 \mathrm{\, M_{\odot}} \leq \mathrm{M_{*}} \leq 8 \mathrm{\, M_{\odot}}$). Mid-infrared (MIR) spectra of oxygen-rich AGB stars show amorphous silicate features at around 10 and $18 \mathrm{\, \mu m}$, which is attributed to Si--O stretching and Si--O--Si bending vibration, respectively. AGB stars contribute significantly to the ISM dust in the Milky Way; their dust injection rate is estimated to be $6\times 10^{-3} \mathrm{\, M_{\odot} \mathrm{\, yr^{-1}}}$, which corresponds to nearly 67\% of the total dust injection in the Galaxy \citep{Tielens2018}. For silicate dust,
 O-rich AGB stars are estimated to account for 29\% of the total silicate dust injection, with a dust injection rate of $5 \mathrm{\, M_{\odot} \mathrm{\, pc^{-2}} \mathrm{\, Myr^{-1}}}$ \citep{Tielens2005}.
  
  Dust grains drive the acceleration of stellar wind via radiation pressure and are ejected into the ISM \citep{Hofner2018}. They regulate the thermal structure of circumstellar environments and the ISM by absorbing UV radiation and emitting IR radiation. Dust also significantly affects material evolution by providing surfaces for chemical reactions and acting as catalysts and reactants in the ISM and molecular clouds \citep{Cabedo2024, Reboussin2014}. Therefore, understanding the physical and chemical properties of dust is important when evaluating the functions of dust in interstellar and circumstellar phenomena.
  
  However, the properties of dust around AGB stars are not fully constrained. The shape of MIR spectra depends on the optical constants of dust, which change with chemical compositions and structures \citep[e.g.,][]{Speck2011, Speck2015}, as well as physical parameters such as particle size \citep{Jaeger1994, Dorschner1995}, temperature \citep{Zeidler2013}, and grain shapes \citep{Fabian2001, Min2005, Min2006}. Observations of dust around oxygen-rich AGB stars provide limited diagnostic information because it is challenging to distinguish individual effects of chemical composition, grain size, dust temperature, and shape from broad spectral features of amorphous silicate.
  
  Moreover, determining where and how iron exists in circumstellar and interstellar environments has been challenging. Iron is one of the most abundant cationic elements, following magnesium and silicon. Thermodynamic equilibrium calculations predict that the stable phase of Fe is metallic iron, assuming the solar composition at high temperatures, although it can form FeS at lower temperatures or Fe-bearing silicate under more oxidizing conditions, and Fe-bearing silicide under more reductive conditions. Theoretical studies suggest that iron tends to initially condense as a metal rather than form silicates in the circumstellar outflow \citep{Nuth1990, Gail1999, Lodders1999}. Even if Fe forms metallic grains, it is crucial to determine whether metallic iron exists inside or separately from amorphous silicate to understand chemical reactions on the dust surfaces in the ISM. The catalytic function of iron changes depending on its chemical state (whether it is metal or an oxide), and on its spatial distribution (whether metal iron exists inside or on the surface of a silicate particle).
  
  Glass with embedded metal and sulfides (GEMS) found in chondritic porous interplanetary dust particles are amorphous silicate grains \citep{Keller2011} that contain metallic iron nanoparticles inside and sulfide particles on the surface \citep{Matsuno2022}. Isotopic analysis of GEMS and experiments forming GEMS analogs suggest that some GEMS originates from AGB stars \citep{Keller2011, Kim2021, Matsuno2021, Enju2022}. \cite{Bradley1999} showed that GEMS grains exhibit IR spectral features similar to those of circumstellar and interstellar amorphous silicates. This spectral match does not necessarily mean that circumstellar dust also has metallic particles. Metallic iron is featureless in the MIR region, and therefore, its effects have been discussed only in terms of near-infrared (NIR) absorption and opacity enhancement.
  
  Since the 1980s, emission and absorption of circumstellar and interstellar dust have been modeled as hypothetical optical constants referred to as “astronomical silicates” \citep{Draine1984, Ossenkopf1992}, which are arbitrarily synthesized by combining the observational data of dust and laboratory measurements. In deriving astronomical silicates, certain optical data of materials containing metallic inclusions are used, such as “dirty silicate” \citep{Jones1976, Rogers1983}, in the case of \cite{Draine1984}, and laboratory silicates [(Mg, Fe)SiO$_{3}$] with Fe and Fe$_{3}$O$_{4}$ inclusions in the case of \cite{Ossenkopf1992}. However, these optical constants do not represent real solid materials and, therefore, cannot be used to discuss the chemical compositions and structures of dust.
  
  In laboratories, experiments have been carried out to produce amorphous silicate dust analogs with various techniques, such as the sol-gel method \citep{Jaeger2003}, the so-called “smoke” method \citep{Nuth2002, Rietmeijer2002}, and melt-quenching \citep{Dorschner1995, Mutschke1998}. Most of these experiments assume that circumstellar dust is pure amorphous silicate without metallic iron inclusions. On the other hand, \cite{Speck2015} proposed the combination of iron-free amorphous silicate with a chondritic composition, (Na$_{0.11}$Ca$_{0.12}$Mg$_{1.86}$)(Al$_{0.18}$Si$_{1.85}$)O$_{6}$, and metallic iron as an alternative to astronomical silicates, assuming that amorphous silicate grains and metallic iron grains exist separately. 
  
 The equilibrium condensation temperature of metallic iron is similar to those of crystalline Mg-silicates (forsterite and enstatite), but kinetically, the condensation of metallic iron can be delayed due to its high surface energy and high efficiency of absorbing stellar light \citep{Kozasa1987, Woitke2006}. The abundant presence of amorphous silicates, which are metastable silicate phases, suggests that circumstellar dust formation generally occurs under nonequilibrium conditions, and the condensation sequence of amorphous silicates and metallic iron is further unclear. Although \cite{Gail1999} estimated that metallic iron first condenses on the surface of silicate, whether the iron exists as separate particles or as inclusions inside the silicate grains has not been determined. \cite{Kemper2002} argued that nonspherical metallic iron grains explain the observation of an AGB star with an optically thick circumstellar dust shell. In the ISM, an estimated 70\% of its depleted iron exists as metallic particles within amorphous silicate, according to numerical simulations  \citep{Zhukovska2018}. While \cite{Jaeger2003} theoretically showed IR spectral features of dust consisting of an amorphous silicate shell and a metallic iron core, the effects of metallic Fe–Ni particles on the IR spectra of amorphous silicate have not been examined experimentally.
  
  Laboratory studies using induction thermal plasma (ITP) systems \citep{Kim2021, Matsuno2021, Enju2022} produced amorphous silicate grains containing metallic iron cores (GEMS-like nanoparticles). The ITP systems can produce nanoparticles for a wider range of chemical compositions than the melt-quenching method by using a high-temperature plasma flame  ($\sim 10^{4} \mathrm{\, K}$)  to vaporize refractory materials and a rapid cooling condition ($\sim 10^{4} - 10^{5}\mathrm{\, K \mathrm{\, s^{-1}}}$). However, the optical constants of amorphous silicate nanoparticles produced with ITP systems, including these GEMS-like nanoparticles, have not been determined. This is because the quantitative measurement of reflectance spectra requires polished samples with smooth and flat surfaces, and it is challenging to derive reliable optical constants of nanopowders. 
  
  The objective of this study was to determine the optical constants of amorphous silicate nanoparticles with a CI chondritic composition, which can be applied to constrain the chemical composition of dust around other AGB stars and to investigate spectral changes with the presence of metallic iron inclusions within amorphous silicate grains. We performed condensation experiments in Mg–Ca–Na–Al–Si–Fe–Ni–O and Mg–Ca–Na–Al–Si–O systems using the ITP system to produce amorphous silicate nanoparticles with a metallic Fe-Ni-containing CI chondritic composition and a Fe-Ni-free CI chondritic composition. We determined the optical constants of nanoparticle products by measuring absorption and reflectance spectra and fitting them using the Lorentz oscillator model. With the derived optical constants, we discuss the spectral changes caused by the presence or absence of metallic iron and the possibility that dust around AGB stars has metallic iron particles.

\section{Methods}
\subsection{Condensation experiments}
  Experiments were conducted in an ITP system (JEOL TP-40020NPS, The University of Tokyo; \citealt{Kim2021}) consisting of a plasma torch, chamber, and powder feeder (JEOL TP-99010FDR). High-temperature plasma flame ($\sim 10^{4}\mathrm{\, K}$) was generated with a fixed input power of $6 \mathrm{\, kW}$. Ar gas with a flow rate of $35 \mathrm{\, L \mathrm{\, min^{-1}}}$ was used as a main plasma-forming gas. The plasma-forming gas was injected vertically outside the quartz tube and generated a long plasma (radial) flame. The chamber pressure was maintained at $\sim 70 \mathrm{\, kPa}$ during the experiments. The starting material was injected into a plasma flame along with an Ar carrier gas at a flow rate of $2 \mathrm{\, L \mathrm{\, min^{-1}}}$ and vaporized with the plasma flame. The gas was quenched by water running outside the chamber ($\sim 10^{4}-10^{5}\mathrm{\, K \mathrm{\, s^{-1}}}$) to condense nanoparticles. The feeding rates of the starting materials were calculated from the total mass of the products and the experimental duration of 15 minutes. The experimental conditions are shown in Table \ref{table:experimental conditions}.

\begin{table*}
\caption{Summary of the experimental conditions.}         
\label{table:experimental conditions}      
\makebox[\textwidth][c]{ %
\begin{tabularx}{0.9\textwidth}{c c c >{\centering\arraybackslash}X >{\centering\arraybackslash}X }
\hline\hline    
Run & $X_O$ & Feeding rate 
    & Plasma forming gas rate (L min$^{-1}$)
    & {Weight of condensates (mg)}\\
  &  & (mg min$^{-1}$) 
  & \makebox[\linewidth]{\hfill Ar \hfill O$_2$ \hfill He \hfill} 
  & \makebox[\linewidth]{\hfill upper \hfill lower \hfill}\\
\hline                    
   CI-1 & 0.93–0.98 & 286 & \makebox[\linewidth]{\hfill 35 \hfill 0 \hfill 0 \hfill} &  \makebox[\linewidth]{\hfill 578.5 \hfill 2175.9 \hfill}\\ 
   CI-2 & 9.35–9.40 & 142 & \makebox[\linewidth]{\hfill 35 \hfill 1 \hfill 0 \hfill} & \makebox[\linewidth]{\hfill 262.4 \hfill 569.9 \hfill}\\
\hline                  
\end{tabularx}
}
\end{table*}

 The starting material was prepared by grinding oxides and metallic power reagents: MgO (>99.99\% purity, $10 \mathrm{\, \mu m}$; Kojundo Chemical Lab. Co., Ltd.), CaCO$_{3}$ (>99\%  purity, $5 \mathrm{\, \mu m}$; Kojundo Chemical Lab. Co., Ltd.), Na$_{2}$SiO$_{3}$ (>99\% purity, prepared by grounding coarse grains; Kojundo Chemical Lab. Co., Ltd.), Al (>99.9\% purity, $3 \mathrm{\, \mu m}$; Kojundo Chemical Lab. Co., Ltd.), Fe (>99.9\% purity, $3-5 \mathrm{\, \mu m}$; Kojundo Chemical Lab. Co., Ltd.), Ni (>99\% purity, $10 \mathrm{\, \mu m}$; Kojundo Chemical Lab. Co., Ltd.), SiO$_{2}$ ($\>99.9\%$ purity, $4 \mathrm{\, \mu m}$; Kojundo Chemical Lab. Co., Ltd.), and Si (>$99$\% purity, $5 \mathrm{\, \mu m}$; Kojundo Chemical Lab. Co., Ltd.). These reagents were mixed at the CI chondritic composition \citep{Lodders2009} in the Mg–Ca–Na–Al–Si–Fe–Ni–O and Mg–Ca–Na–Al–Si–O systems (Table \ref{table:SM}). 
 
\begin{table*}
\caption{Chemical composition of starting materials.}      
\label{table:SM}      
\centering             
\begin{tabular}{c c c c c c c c c c}          
\hline\hline                        
Run & Chemical composition of & MgO & CaCO$_{3}$ & Na$_{2}$SiO$_{3}$ & Al & SiO$_{2}$ & Si & Fe & Ni\\ 
 & starting material &  \multicolumn{8}{c}{(mol \%)} \\   
\hline                                  
    CI-1 & Mg$_{1.03}$Ca$_{0.06}$Na$_{0.06}$Al$_{0.08}$SiFe$_{0.85}$Ni$_{0.05}$ & 1.03 & 0.06 & 0.06 & 0.08 & 0.95 & 0.05 & 0.85 & 0.05\\    
    CI-2 & Mg$_{1.03}$Ca$_{0.06}$Na$_{0.06}$Al$_{0.08}$Si & 1.03 & 0.06 & 0.06 & 0.08 & 0.97 & ... & ... & ... \\ 
\hline                                          
\end{tabular}
\end{table*}

 The redox condition in the ITP system was estimated by using the total oxygen abundance in the system relative to the oxygen abundance to convert cations other than Fe and Ni into oxides, $X_{\mathrm{O}}$ \citep{Enju2022}:
    \begin{equation} \label{eq:Xo}
     X_{\mathrm{O}} =\frac{\left(A_{\mathrm{O, SM}}+ {R_\mathrm{O}}/{f}\right)}{A_\mathrm{O}}                        
    .\end{equation}
Here $f$ is the feeding rate of the starting materials, $R_\mathrm{O}/{f}$ is the oxygen inflow rate into the ITP system, which is estimated to be $0.5-3.6\times 10^{-4} \mathrm{\, mol \mathrm{\, min^{-1}}}$ by \cite{Enju2022}, $A_{\mathrm{O, SM}} $ is the oxygen abundance of the starting material, and Ao is the oxygen abundance of the starting material, assuming that all Si and Al are fully oxidized. Fe condenses as a metal under reducing conditions at $X_{\mathrm{O}} < 1$, and is incorporated into silicates as a cation under oxidizing conditions at $X_{\mathrm{O}} > 1$. In the Fe-Ni-containing system, we aimed to obtain amorphous silicates with Fe-Ni metal inclusions by controlling $X_{\mathrm{O}}$ with the range of 0.93-0.98, by setting the SiO$_2$\,/\,(Si~+~SiO$_2$) ratio to 0.95 in the starting materials. In the Fe-Ni-free system, $X_{\mathrm{O}}$ was set to be 9.35-9.40. All Si was introduced as Si powders instead of SiO$_2$, and O$_2$ plasma-forming gas was added at a flow rate of $1 \mathrm{\, L \mathrm{\, min^{-1}}}$ to prevent the formation of any metallic particles and silicides.
 
 The condensates were collected from the upper and lower walls of the chamber separately. With a radial flame, the size distribution of the condensates does not differ with the distance from the plasma flame in a chamber. The bulk chemical composition and peak position in absorption spectra of the products from the upper and lower walls were almost identical. As more condensates were collected from the lower wall, we used the lower-wall product for analyses.

\subsection{Analytical methods}
 The crystal phases of the condensates were examined by powder X-ray diffraction (XRD; Rigaku RINT-2100, the University of Tokyo) using Ni-filtered Cu K$\alpha$ radiation ($\lambda = 1.5418 \mathrm{\,\AA}$) at an accelerating voltage of $40 \mathrm{\, kV}$ and a tube current of $30 \mathrm{\, mA}$. A zero-diffraction sample holder of Si single crystal was used. The XRD patterns were measured with fluorescent X-ray reduction mode. The scan range was $15-85 \mathrm{\, ^\circ}$ in $2 \theta$, and the scan speed was $0.4^\circ \mathrm{\, min^{-1}}$.
 
 The bulk chemical composition of the products was measured using an electron probe microanalyzer (EPMA; JEOL JXA-8530F, the University of Tokyo). The condensates were pressed onto Cu plates at $15 \mathrm{\,MPa}$ for 1~minute using an oil hydraulic press and subsequently coated with carbon. The pressed samples were analyzed with a defocused electron beam of $2 \mathrm{\,\mu m}$ in diameter at an accelerating voltage of $12 \mathrm{\, kV}$ and a beam current of $12 \mathrm{\, nA}$. Approximately 20 points were selected from the homogeneous sample surface regions without large particles ($>1  \mathrm{\,\mu m}$ in diameter), which were supposed to be residues of the starting materials. Albite (NaAlSi$_3$O$_8$), periclase (MgO), Al$_2$O$_3$, wollastonite (CaSiO$_3$), Fe$_2$O$_3$, and NiO were used as standard materials. The measured data were corrected using the ZAF (atomic number, absorption, and fluorescence) correction.
 
 The structures of the condensates were observed and the chemical composition of the nanoparticles was measured using transmission electron microscopy (TEM; JEOL JEM-2800, the University of Tokyo) equipped with energy dispersive X-ray spectroscopy (EDS). The condensed particles were mounted onto Cu TEM grids coated with carbon without any dispersant. Dark field scanning transmission electron microscopy      (STEM) images and STEM-EDS maps with a resolution of $256 \times 256$ pixels were obtained. To minimize beam damage to the sample, a beam with an accelerating voltage of  $200 \mathrm{\, kV}$ and a spot size of $0.2 \mathrm{\, nm}$ was used for a one-hour measurement.
 
 The products were analyzed with Fourier transform infrared spectroscopy (FT-IR) to obtain transmittance and reflectance spectra. For measuring transmittance spectra, the condensates were dispersed in KBr powder at a mass ratio of 1:500, evacuated for 10 minutes, and pressed at $30 \mathrm{\,MPa}$ for 10 minutes into $10 \mathrm{\, mm}$ diameter disks using an oil hydraulic press under vacuum. The transmittance spectra of the sample pellets were obtained over the range of $1.28-25.6 \mathrm{\,\mu m}$ with a resolution of $2 \mathrm{\, cm}^{-1}$ (JASCO FT/IR-4200, The University of Tokyo).
 
 For reflectance measurements, only the condensed nanopowders were pressed into pellets. In addition to a gold mirror, which is commonly used as a reference for reflectance measurements, we prepared four reference materials to evaluate the scattering effects caused by the surface roughness of the nanoparticle-made pellets. The CI-1 and CI-2 pellets were coated with gold of approximately $50 \mathrm{\, nm}$ and  $100 \mathrm{\, nm}$ thickness using a sputter coater (HITACHI sputter coater E-1030, The University of Tokyo). Although the CI-1 and CI-2 pellets were coated with gold simultaneously, a slight difference in their distances from the evaporation source resulted in a thinner Au coating on CI-1 than on CI-2. Reflectance spectra were measured at an incident light angle of $10 \mathrm{\, ^{\circ}}$ over the MIR range of $375-7400 \mathrm{\, cm}^{-1}$ with a resolution of $4 \mathrm{\, cm}^{-1}$ (Thermo NICOLET6700, Osaka Sangyo University).

\section{Results}
\subsection{Bulk analysis (XRD and EPMA)}
\begin{figure}
\centering
   \includegraphics[width=\hsize]{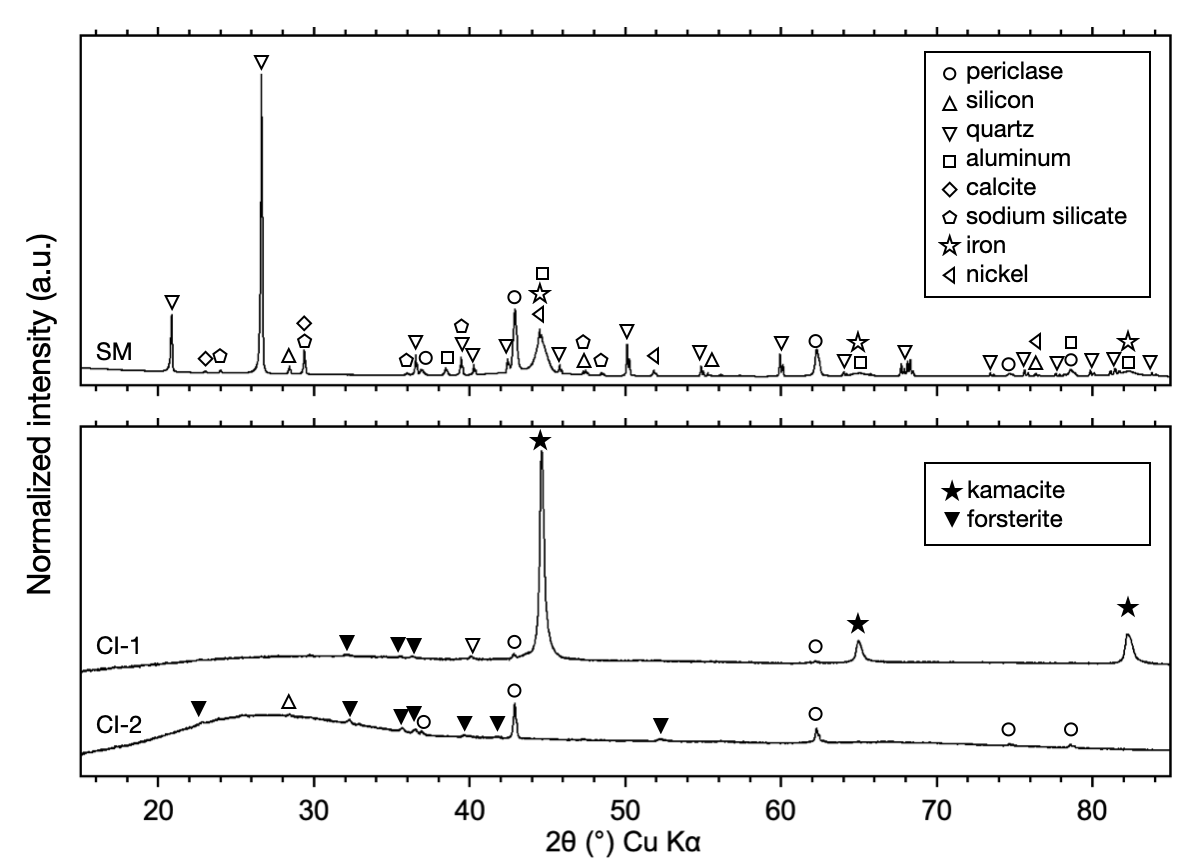}
     \caption{XRD patterns of the starting material of CI-1 (upper panel) and experimental products (CI-1, CI-2; bottom panel). Peaks indicated by open symbols are derived from the reagents composing the starting material. Filled symbols show the peaks of new phases formed through the condensation.}
     \label{fig:XRD}
\end{figure}

 Figure \ref{fig:XRD} shows the XRD patterns of the products of both runs and the starting material of Run CI-1. Almost all of the starting materials were vaporized during the experiments. The condensates of both runs showed broad halos in the range $2\theta = 20-40 \mathrm{\, ^{\circ}}$, indicating the presence of amorphous materials, along with a few peaks derived from residual starting materials. Strong peaks corresponding to kamacite were observed in the product of Run CI-1. Additionally, crystalline peaks of periclase (MgO), silicon, and quartz (SiO$_2$) were detected, which originated from the unvaporized reagents in the starting materials. Forsterite (Mg$_2$SiO$_4$) is supposed to be newly crystallized from the unvaporized melt, as discussed in \cite{Kim2021}.  The weak peak intensities of the crystalline phases indicate that the phases from unvaporized reagents are present in negligible amounts compared with the amorphous silicates.

 Table \ref{table:EPMA} shows the bulk chemical composition of products determined using an EPMA and the initial compositions of their starting materials. The products exhibit lower content for some elements (Mg, Ca, Na, and Al), and higher Fe and Ni contents compared to the initial compositions. This discrepancy occurred because the starting materials were not fully vaporized, as well as the surface roughness and the high porosity of the pressed samples reduced the total weight percentage measured by an EPMA. The Na content was especially low (to approximately 17\% of the initial value), because Na dissipated as a gas phase during the experiments due to its high volatility.

\begin{table}
\caption{Bulk compositions of products (at.\%) determined using an EPMA compared with the initial composition of starting materials.}      
\label{table:EPMA}      
\centering          
\begin{tabular}{c c c c c}      
\hline\hline         
 & \multicolumn{2}{c}{CI-1} & \multicolumn{2}{c}{CI-2}\\  
 & {Starting} & \multirow{2}{*}{Product} & {Starting}  & \multirow{2}{*}{Product}\\ 
 & {material} &   & {material}  &  \\  

\hline                                   
    Mg & 1.03 & 0.84 & 1.03 & 0.92 \\ 
    Ca & 0.06 & 0.03 & 0.06 & 0.03 \\ 
    Na & 0.06 & 0.01 & 0.06 & 0.01 \\ 
    Al & 0.08 & 0.05 & 0.08 & 0.07 \\ 
    Si & 1.00 & 1.00 & 1.00 & 1.00 \\ 
    Fe & 0.85 & 0.91 & ... & ... \\ 
    Ni & 0.05 & 0.06 & ... & ... \\ 
    O & 3.40 & 2.9 & 3.44 & 2.9 \\ 
\hline                                             
\end{tabular}
\end{table}
 
\subsection{TEM observation}
\begin{figure}
\centering
   \includegraphics[width=\hsize]{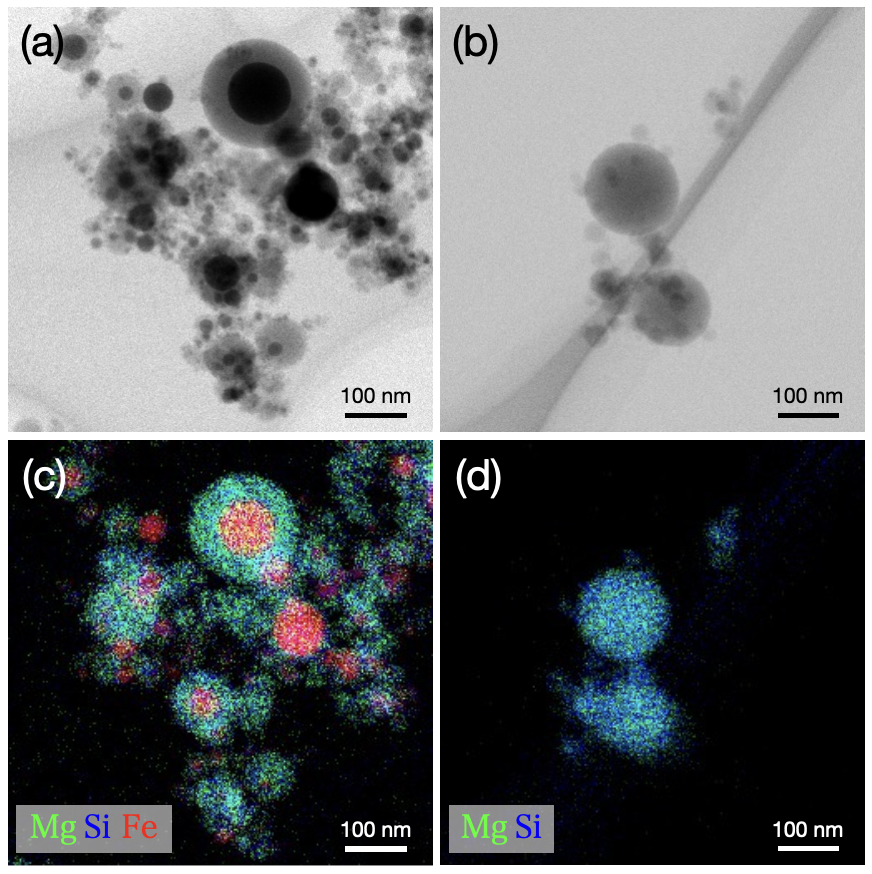}
     \caption{TEM bright field images of experimental products in CI-1 (a) and CI-2 (b), and STEM-EDS maps of experimental products in CI-1(c) and CI-2 (d).}
     \label{fig:TEM}
\end{figure}

\begin{figure}
\centering
   \includegraphics[width=\hsize]{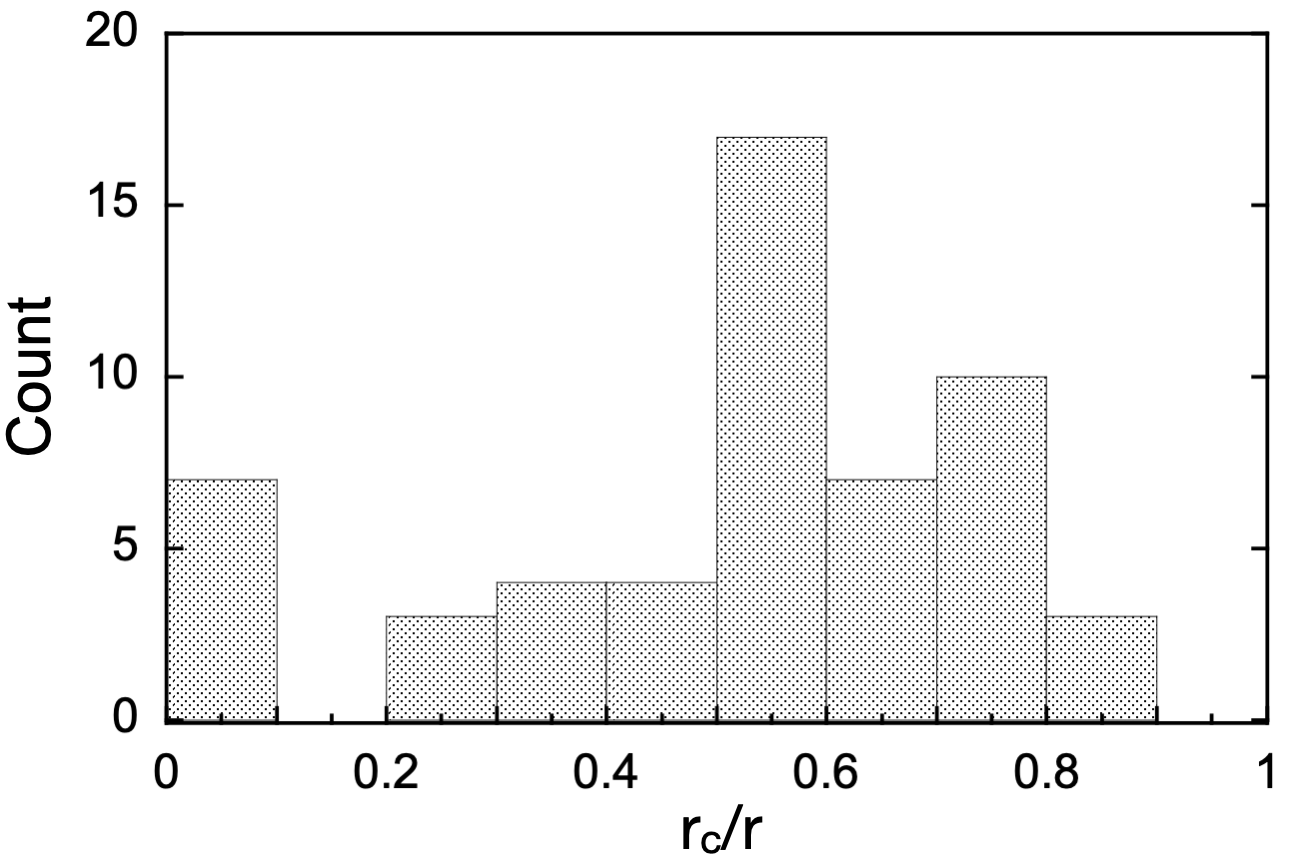}
     \caption{Histogram of the core size ratios, $r_{\rm{c}}/r$, of particles condensed in CI-1.}
     \label{fig:rc}
\end{figure}

 Figure \ref{fig:TEM} presents bright field TEM images and STEM-EDS maps of the products. The condensates were spherical amorphous silicate grains with a size of approximately 10 to $200 \mathrm{\, nm}$ in diameter. In CI-1, kamacite particles ($\sim 10-100 \mathrm{\, nm}$ in diameter) were also observed. These metallic particles were supposed to be embedded inside amorphous silicate grains, as confirmed with 3D STEM-EDS tomography in the previous studies \citep{Matsuno2021, Kim2021}. The diameter ratio of a kamacite particle and whole silicate grain, $r_{\rm{c}}/r$, varied from 0 to 0.87 (Fig. \ref{fig:rc}), and the average value was approximately 0.50. Note that not all particles had a metallic core and core size ratios $r_{\rm{c}}/r$ were only examined for particles larger than $\sim 50 \mathrm{\, nm}$ in diameter.
 
 The chemical compositions of amorphous grains and kamacite particles were measured with STEM-EDS. The STEM beam conditions were optimized to avoid beam damage on the amorphous silicate grains. Only maps showing no significant temporal change of textures and elemental compositions were used to derive the chemical compositions. In CI-2, the Na count was below the qualification limit. The representative chemical compositions of amorphous silicate particles were determined from the data with the highest counts. The average Mg/Si ratio of amorphous silicate and the Fe/(Fe+Ni) ratio in kamacite are shown in Table \ref{table:STEM}. The Mg/Si ratios were homogeneous among the grains and consistent between the two runs. The relative depletion of Mg content accounts for the residue of periclase (MgO) in starting materials as seen in XRD patterns (Fig. \ref{fig:XRD}). In CI-1, not all Fe in the starting material condensed as kamacite, but a small fraction of Fe was incorporated into the amorphous silicate structure as Fe$^{2+}$, modifying the SiO$_4$ networks of amorphous silicate with Fe/(Fe+Mg) $\sim$ 0.08.

\begin{table*}
\caption{Chemical compositions of products measured with STEM-EDS.}     
\label{table:STEM}      
\centering          
\begin{tabular}{c c c c c c c c c c}      
\hline\hline         
 & \multicolumn{7}{c}{Chemical composition of amorphous silicate} & Mg/Si in amorphous silicate & Fe/(Fe+Ni) in kamacite\\  
 & Mg & Ca & Na & Al & Si & Fe & Ni & (Average $\pm$ S.D.) & (Average $\pm$ S.D.)\\ 
\hline   
    CI-1 & 1.05 & 0.03 & 0.01 & 0.06 & 1.00 & 0.09 & 0.00 & $0.97 \pm 0.09$ & $0.93 \pm 0.07$ \\ 
    CI-2 & 0.96 & 0.06 & 0.00 & 0.07 & 1.00 & ... & ... & $0.98 \pm 0.10$ & ... \\ 
\hline                                             
\end{tabular}
\end{table*}

\subsection{Infrared spectroscopy}
\begin{figure}
\centering
   \includegraphics[width=\hsize]{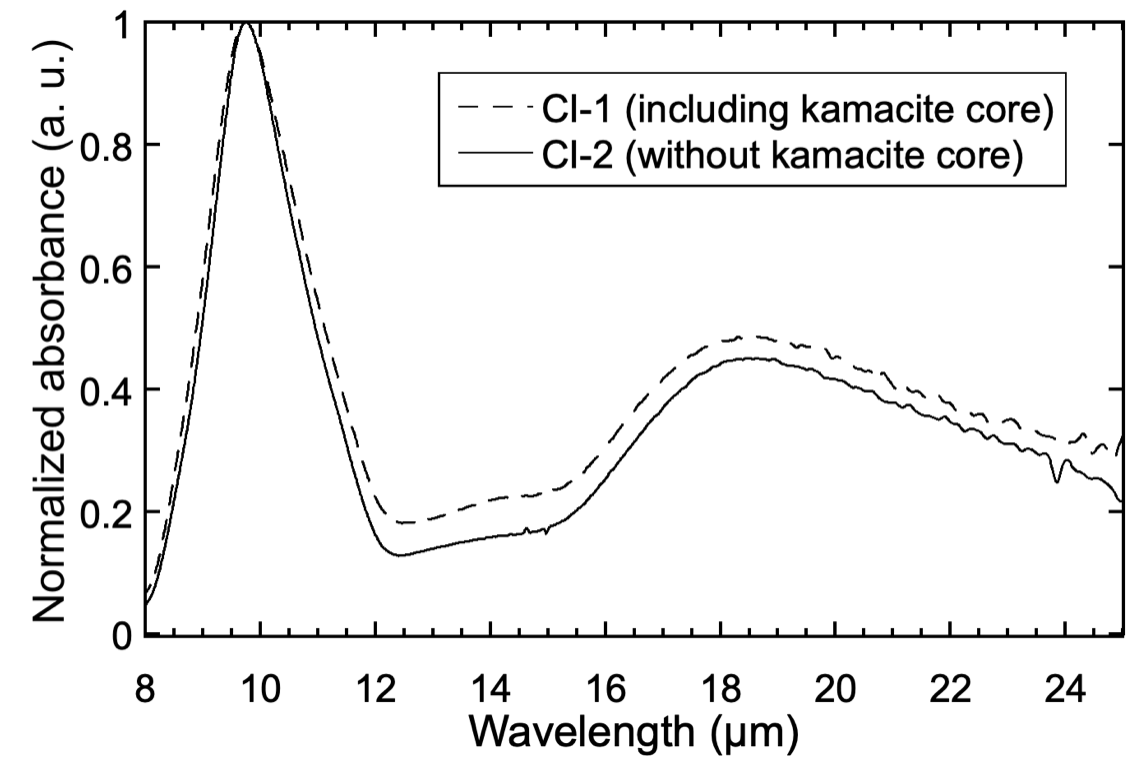}
     \caption{FT-IR absorbance spectra of CI-1 (dashed line) and CI-2 (solid line).}
     \label{fig:Abs}
\end{figure}

 The IR absorbance spectra of experimental products show broad peaks at approximately $10 \mathrm{\, \mu m}$ and $18 \mathrm{\, \mu m}$ (Fig. \ref{fig:Abs}). These peaks are derived from Si--O stretching and Si--O--Si bending vibrations in amorphous silicates, respectively. There was no difference in the positions of the 9.8 and $18.5 \mathrm{\, \mu m}$ peaks between CI-1 and CI-2. 
 
\begin{figure}
\centering
   \includegraphics[width=\hsize]{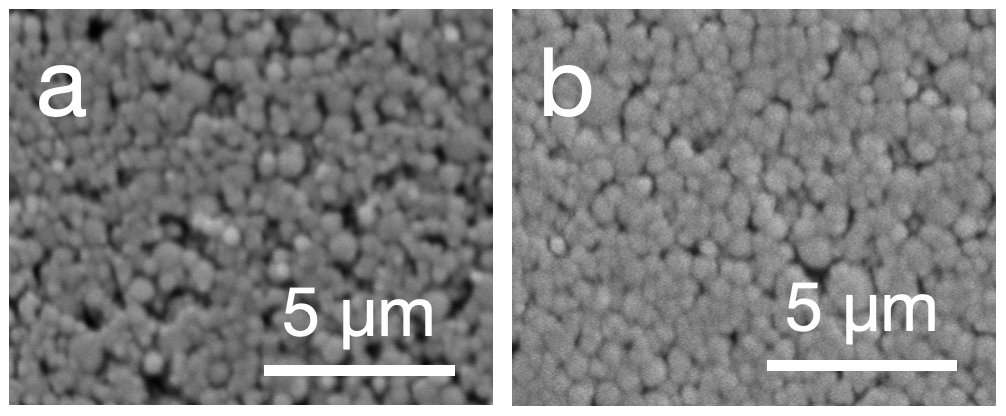}
     \caption{Secondary electron images of the surface of a thin (a)  and a thick (b)  Au-coated pellet of CI-2 taken with a $0.4 \mathrm{\, nA}$ electric beam at $5 \mathrm{\, kV}$ (JEOL; JSM-7000F).}
     \label{fig:SEM}
\end{figure}

 For IR reflectance measurements of polished samples such as quenched glass, a gold mirror is generally used as a reference. However, unlike polished samples, the pressed pellets of nanopowders used in this study are not as flat as the gold mirror.  To account for the decrease in reflectance caused by surface roughness, pressed pellets coated with thin ($\sim 50 \mathrm{\, nm}$) and thick ($\sim 100 \mathrm{\, nm}$) gold films were prepared. The thin ($\sim 50 \mathrm{\, nm}$) gold coating did not alter the original surface roughness due to the agglomerated condensed nanoparticles, while the thick ($\sim 100 \mathrm{\, nm}$) coating modified the surface by filling in the roughness (Fig. \ref{fig:SEM}).
 
\begin{figure}
\centering
   \includegraphics[width=\hsize]{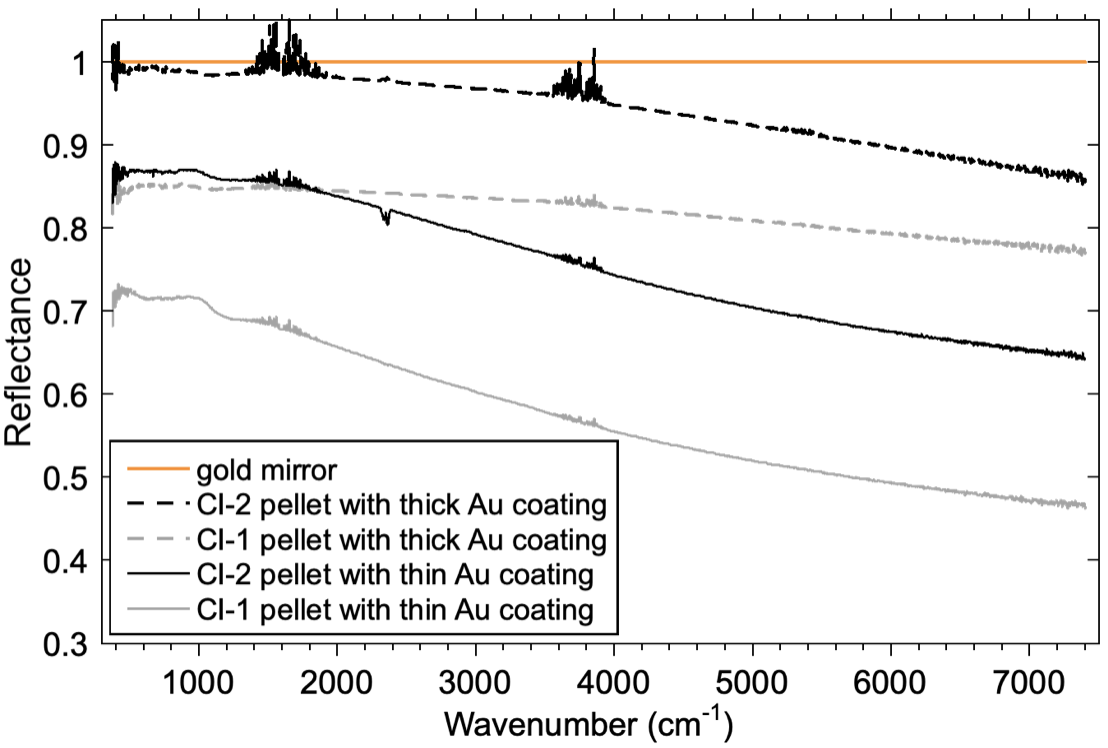}
     \caption{Reflectance of Au-coated samples measured with a gold mirror. The reflectance of a thin Au-coated pellet, a thick Au-coated pellet, and a gold mirror were divided by the reflectance of a gold mirror. CI-1 pellets have a slightly thinner coating overall than CI-2 pellets because of their positions during deposition.}
     \label{fig:reference}
\end{figure}

 To evaluate reference materials for reflectance measurements, the reflectance of each reference was measured. Figure \ref{fig:reference} compares the IR reflectance spectra of five reference materials (thin Au-coated pellets of CI-1 and CI-2, thick Au-coated pellets of CI-1 and CI-2, and gold mirror) divided by the reflectance of the gold mirror. 

 In both the CI-1 and CI-2 runs, the reflectance near $400 \mathrm{\, {cm}^{-1}}$ was approximately 0.15 larger for the thick Au-coated pellets than for the thin Au-coated pellets. This difference indicates that reflectance decreases with increasing porosity on the pellet surface. The surface roughness was preserved on the thin Au-coated pellets better than the thick Au-coated pellets, as shown in Fig. \ref{fig:SEM}. Also, the reflectance of thick and thin Au-coated pellets of CI-2 was systematically larger than that of CI-1. It is inferred that materials under the Au-coating affect the reflectance.
 
 The reflectance of Au-coated pellets decreased with increasing wavenumber because gold coated on the rough surface strongly scattered the light, which was not observed on the pellets without gold-coating. The decrease was greater in thin Au-coated pellets than in thick Au-coated pellets (Fig. \ref{fig:reference}). Thick Au-coating reduced the surface roughness of the pellets.
 
 The reflectance spectra of thin Au-coated pellets of CI-1 showed a broad peak at $1000 \mathrm{\, {cm}^{-1}}$, which is derived from amorphous silicate. This indicates that the MIR light penetrated the Au-coating and was reflected by the underlying amorphous silicate particles during the measurement. 
 
 We adopted the thin Au-coated pellet of CI-2 as a reference material, because it preserves the surface roughness of pressed nanoparticles and can correct the decrease in reflectance caused by the surface porosity. Furthermore, no significant peak of amorphous silicate was observed at $1000 \mathrm{\, {cm}^{-1}}$.
 
\begin{figure}
\centering
   \includegraphics[width=\hsize]{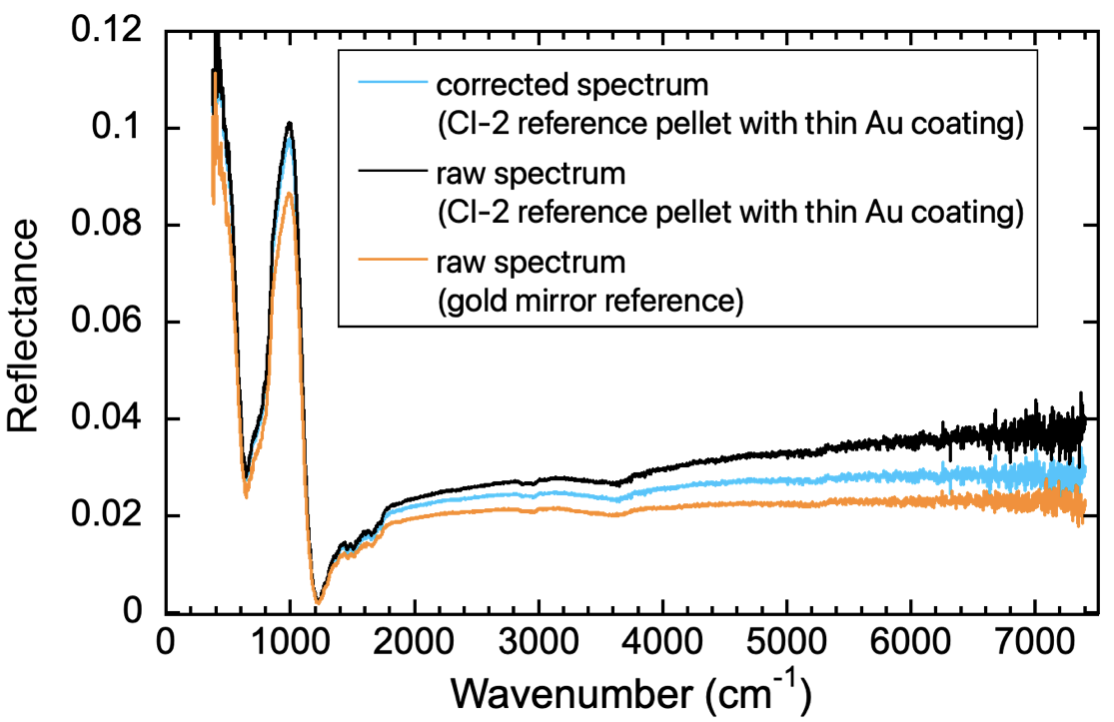}
     \caption{Reflectance spectra of the product (CI-2) measured using a thin Au-coated pellet of CI-2 before (black line) and after correcting for the effects of scattering (blue line), and the reflectance spectrum measured using a gold mirror (orange line).}
     \label{fig:raw_corrected_mirror}
\end{figure}

 Subsequently, we corrected the scattering effect of the thin Au-coated pellet of CI-2. The reflectance spectrum of the thin Au-coated pellet of CI-2 was approximated by a linear function of the wave number. The correction function is obtained so that the slope of the approximation line equals 1, the same as the spectrum of the gold mirror. Using the correction function, the reflectance spectra of products measured with a thin Au-coated pellet of CI-2 were modified to parallel the spectra using a gold mirror in the high wavenumber region. Figure \ref{fig:raw_corrected_mirror} shows the result of the correction, comparing the reflectance spectra of the iron-free product (CI-2) measured using the thin Au-coated pellet before and after the correction, and the spectrum measured using a gold mirror. 
 
\begin{figure}
\centering
   \includegraphics[width=\hsize]{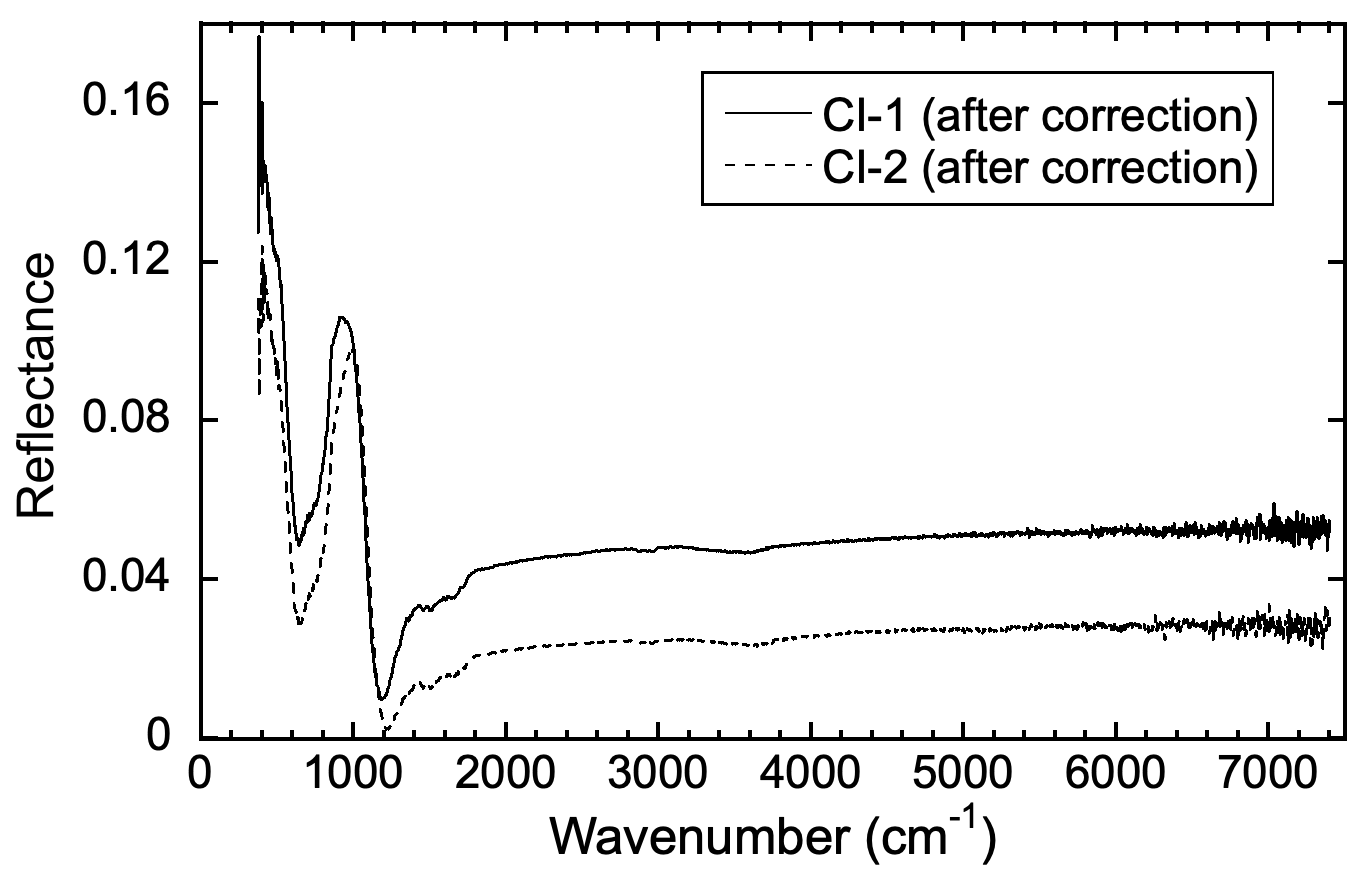}
     \caption{Reflectance spectra measured with a thin Au-coated pellet after correcting for the effects of scattering of CI-1 (solid line) and CI-2 (dashed line).}
     \label{fig:corrected_Ref}
\end{figure}

 The reflectance spectra after correcting the scattering effects of CI-1 and CI-2 are shown in Fig. \ref{fig:corrected_Ref}. Amorphous silicate features at approximately 1000 and $550\mathrm{\, {cm}^{-1}}$ corresponding to wavelengths of $\sim$10 and $18 \mathrm{\, \mu m}$, appeared in both spectra.
 
\subsection{Determining optical constants of experimental products}
 The absorbance spectra of the products dispersed in a KBr pellet are not appropriate to compare with the observed dust emission because the KBr medium effects change the peak positions of the absorbance spectra \citep{Bohren1983,Tamanai2006}.  The optical constants of the amorphous silicate nanoparticles condensed in the run CI-2 were determined from IR absorbance and reflectance spectra by using the Lorentz oscillator model \citep{Bohren1983}.  The complex refractive index $N = n +ik$ is expressed as a function of wavenumber, $\omega$ (Eq. \ref{eq:N2}):
    \begin{equation} \label{eq:N2}
    N^2 = (n + i k)^2 = \sum_j \frac{\omega_{\mathrm{p}j}^2}{\omega_j^2 - \omega^2 - i \gamma_j \omega} + \epsilon_\infty
    .\end{equation}
 Here, $\omega_{\mathrm{p}j}$, $\omega_j$, and $\gamma_j$ represent the resonance frequency, resonance strength, and damping coefficient of the $j$-th oscillator, respectively. $\epsilon_\infty$ is the dielectric constant at high wavenumbers.  $\epsilon_\infty$ was determined to be 1.95 by fitting the reflectance spectra at wavenumbers above $\sim 4000 \mathrm{\, cm^{-1}}$using Eq. \ref{eq:R} as follows:
    \begin{equation}\label{eq:R}
    R = \left| \frac{ \sqrt{(n + i k)^2 - \sin^2\theta} - \cos\theta }{ \sqrt{(n + i k)^2 - \sin^2\theta} + \cos\theta } \right|^2
    ,\end{equation}
 where $\theta$ is angle of incident light in the measurement ($\theta = 10^\circ$).  We note that the reflectance calculated using the optical constants derived from the absorbance spectrum, as below, did not match the measured one at the wavenumbers shorter than $\sim2000 \mathrm{\, cm^{-1}}$. This is probably due to absorption by the deformed or aggregated particles on the surface of the pressed pellet, while particles for absorbance measurements were spherical grains well dispersed in KBr. The reflectance spectra at high wavenumbers were used to determine $\epsilon_\infty$ because there is no molecular vibration resonances.

\begin{figure}
\centering
   \includegraphics[width=\hsize]{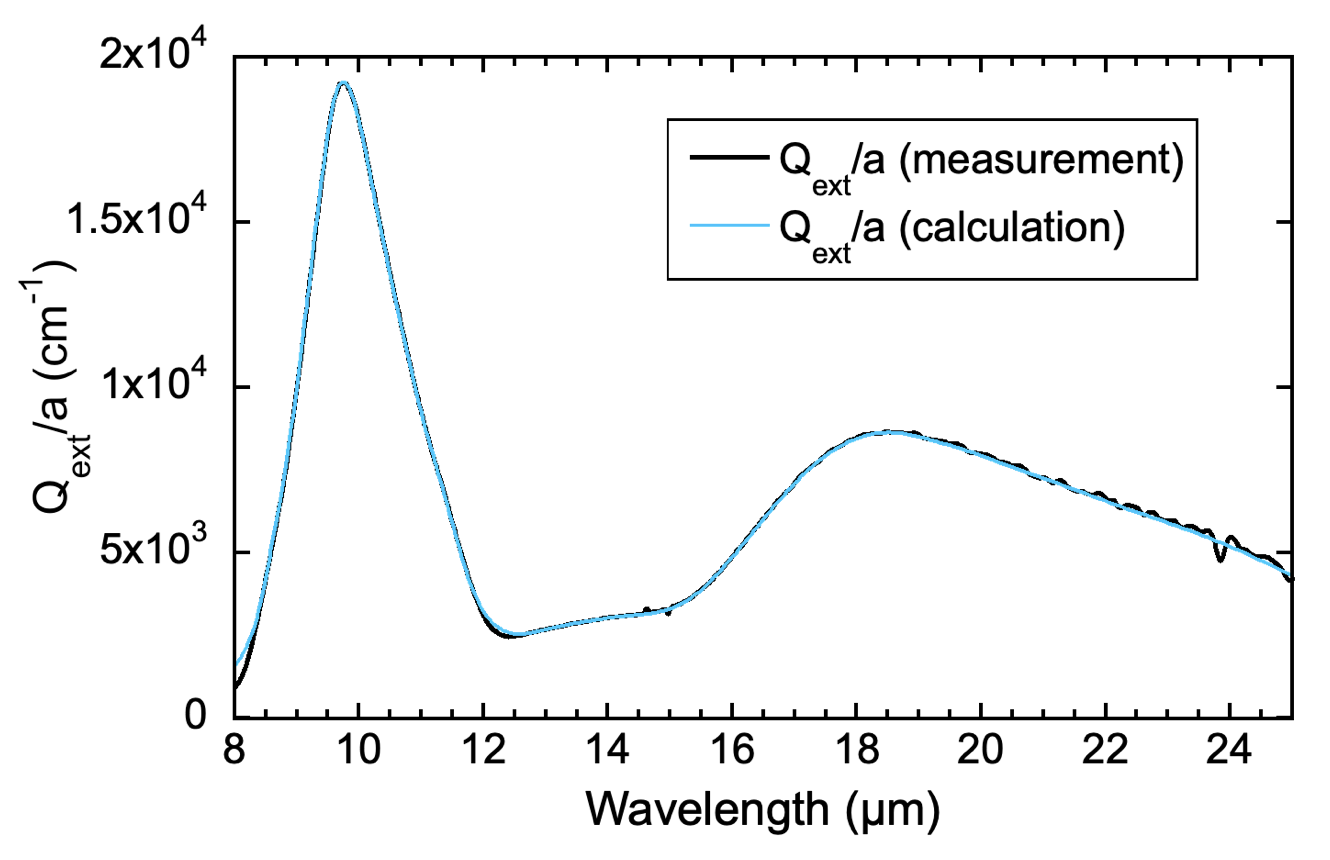}
     \caption{ ${Q_{\mathrm{ext}}^{\mathrm{meas}}}/a$ values obtained via  Eq. \ref{eq:Qext_m} (black line) and via Eq. \ref{eq:Qext} (blue line).}
     \label{fig:Qameas_Qacalc}
\end{figure}

The extinction coefficient ($Q_{\mathrm{ext}}^{\mathrm{meas}}$) of the product was calculated from the measured transmittance A as follows: \begin{equation}\label{eq:Qext_m}
    \frac{Q_{\mathrm{ext}}^{\mathrm{meas}}}{a} = \frac{4\rho S}{3M} \left( \ln{10} \right) A .\end{equation}
 Here $a$ is the average radius of the particle that satisfies ${2\pi a}/{\lambda}\ll1$ as observed with TEM (Fig. \ref{fig:TEM}), $\rho$ is the average density of amorphous silicate, M is the average mass of amorphous silicate in the pellet ($4.0 \times 10^{-4} \mathrm{\,g}$) and S is the surface area of the pellet ($100 \mathrm{\, mm^{2}}$). The average density $\rho$ was measured with a gas displacement pycnometer (AccuPycII 1340; Shimadzu Techno-Research, Inc.) to be $\rho = 2.79 \mathrm{\,g \, cm^{-3}}$. In contrast to the reflectance measurement, the absorbance spectra include the KBr medium effects. Generally, there is a large uncertainty in optical constants determined from absorbance due to grain shape. All condensates in this study, however, were spherical in shape as confirmed in TEM images. Therefore, extinction coefficient ($Q_{\mathrm{ext}}$) of particles in a medium can be expressed with $n$ and $k$ as follows assuming the spherical grain shape \citep{Koike1995}:
    \begin{equation} \label{eq:Qext}
    \frac{Q_{\mathrm{ext}}}{a} = \frac{8\pi}{\lambda} n_0 \, \mathrm{Im} \, \frac{(({n + ik})/{n_0})^2 - 1}{(({n + ik})/{n_0})^2 + 2}
    ,\end{equation}
 where $n_{\mathrm{0}}$ represents the refractive index of the medium ($ n_{\mathrm{0}} = 1.56 $ in KBr). The oscillator parameters ($\omega_{pj}$, $\omega_j$, and $\gamma_j$) were changed until the $Q_{\mathrm{ext}}/a$ calculated by Eq. \ref{eq:Qext} fits the $Q_{\mathrm{ext}}^{\mathrm{meas}}/a$ calculated from the measured absorbance spectrum in the range of $8\mathrm{-}25 \mathrm{\, \mu m}$ as shown in Fig. \ref{fig:Qameas_Qacalc}. 
 
\begin{figure}
\centering
   \includegraphics[width=\hsize]{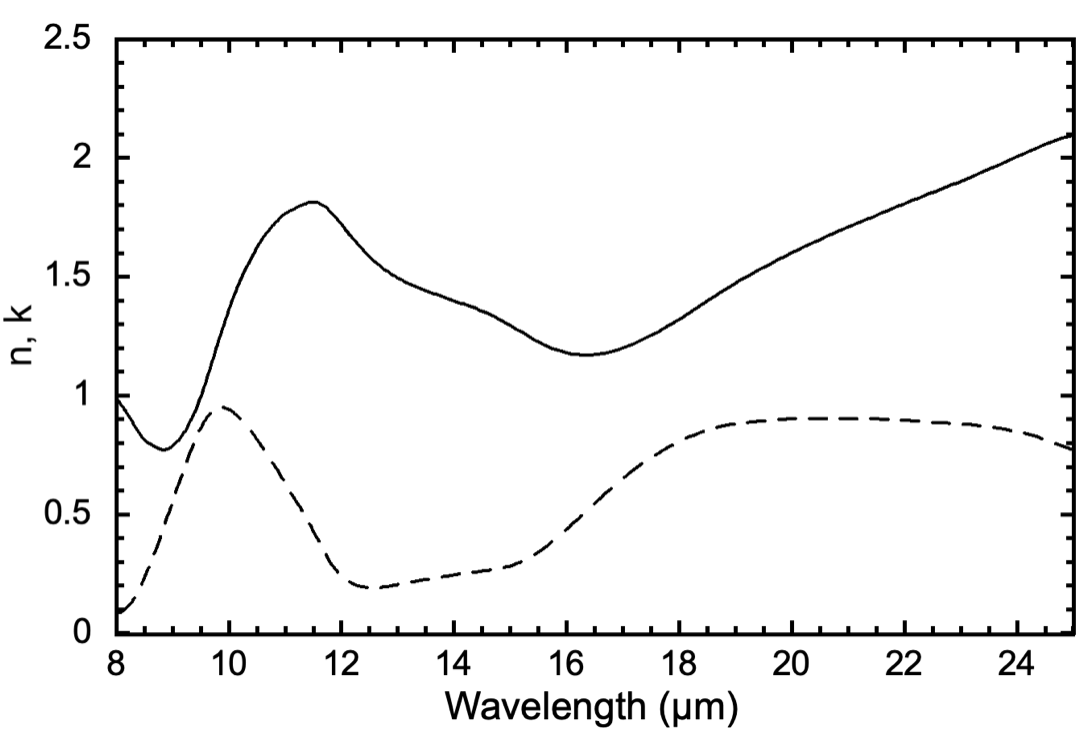}
     \caption{Optical constants ($n$ and $k$) of the amorphous silicate obtained via fitting based on the Lorentz oscillator model. The solid line shows $n,$ and the dashed line shows $k$.}
     \label{fig:nk}
\end{figure}

\begin{figure}
\centering
   \includegraphics[width=\hsize]{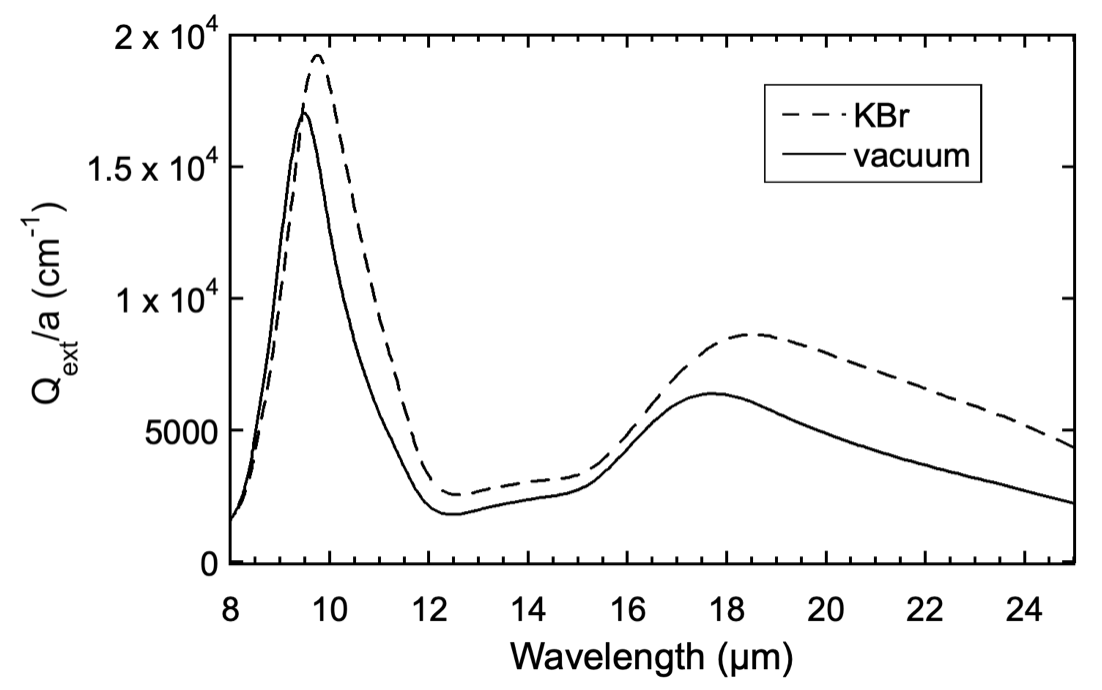}
     \caption{Comparison of $Q_{\mathrm{ext}}/a$ for the CI-2 product, assuming KBr (dashed line) and a vacuum (solid line) as the medium.}
     \label{fig:KBrvsVac}
\end{figure}

 The optical constants $n$ and $k$, which were determined by fitting the absorbance spectra at $8\mathrm{-}25 \mathrm{\, \mu m}$ using $\epsilon_\infty = 1.95$, are shown in Fig. \ref{fig:nk} and Table \ref{table:nk}. Using these optical constants, we calculated the absorbance spectra of spherical grains in vacuum to compare with the observations of circumstellar dust assuming $n_\mathrm{0}=1$, the refractive index of vacuum (Fig. \ref{fig:KBrvsVac}). The peaks become weaker and shift to shorter wavelengths in vacuum than in the KBr medium.

\section{Discussion}
\subsection{Evaluation of a new method for determining optical constants of amorphous silicate nanoparticles}
Previously, optical constants of amorphous silicate dust analogs have been mainly obtained by measuring the reflectance of silicate glasses synthesized with the sol-gel method \citep{Jaeger2003} or melt-quenching \citep{Dorschner1995, Mutschke1998, Speck2015}. This approach allows the derivation of optical constants without relying on uncertain assumptions, as the reflectance spectra of glasses are straightforward to measure. In contrast, measuring the reflectance of amorphous silicate nanoparticles has been challenging due to scattering effects caused by the surface roughness of pelletized powders. Although alternative methods for deriving optical constants without reflectance measurements have been proposed \citep{Koike1989, Koike1995},  $\epsilon_\infty$ was treated as a free fitting parameter or assumed \citep{Koike1989, Koike1995,Takigawa2019}. 
 In this study, we measured both reflectance and absorbance spectra to derive the optical constants of condensed amorphous silicate nanoparticles. Ideally, optical constants derived solely from absorbance would also account for the measured reflectance and vice versa. However, as discussed in Sect. 3.4, the optical constants determined by independently fitting the absorbance spectra were not consistent with measured reflectance, probably because grain deformation of pressed pellets changed silicate features at $500\mathrm{-}1000\mathrm{\, cm^{-1}}$. Consequently, we used the reflectance spectra at the limited range of high wavenumbers $>\sim 4000\mathrm{\, cm^{-1}}$ to determine $\epsilon_\infty$ and derived optical constants by fitting the absorbance spectra based on the Lorentz oscillator model. This approach is reliable for grains with known simple shapes, as the perfectly spherical grains observed in this study (Fig. \ref{fig:TEM}) allow the use of Mie theory without additional shape assumptions.
 Compared to the previous approach to deriving optical constants of nonspherical grains without reflectance measurements \citep{Koike1989, Koike1995}, our method appears more quantitative.  
 
 The reliability of determining $\epsilon_\infty$ from reflectance measured in this study requires further discussion.  The reflectance of CI-2 (Ca$_{0.05}$Mg$_{0.92}$Na$_{0.06}$Al$_{0.07}$Si) at high wavenumbers $> 4000\mathrm{\, cm^{-1}}$ was $\sim $0.029. However, reflectance of silicate glasses with similar Mg/Si ratios synthesized in previous studies showed different values at high wavenumbers. The reflectance at $4000-10000 \mathrm{\, cm^{-1}}$ of silicate glass (MgSiO$_{3}$) produced with the sol-gel method \cite{Jaeger2003} and melt-quenching \cite{Dorschner1995} was $\sim0.045$. 
 
The reflectance at high wavenumbers of our samples with a pyroxene-like Mg/Si ratio (Ca$_{0.05}$Mg$_{0.92}$Na$_{0.06}$Al$_{0.07}$Si) has a smaller value of $\sim$0.029 than the $\sim$0.045 of MgSiO$_3$ glass made with sol-gel method \citep{Jaeger2003} and melt-quenching \citep{Dorschner1995}. Structures such as porosity and density can differ depending on the synthesis method \citep{Speck2011}. These structural differences among silicate glasses and amorphous silicate particles may lead to different reflectance at high wavenumbers. Different types of samples used to measure the reflectance can cause variations in the value of reflectance. While the samples in this study and \cite{Jaeger2003}  are powders, the quenched glass in \cite{Dorschner1995} is the bulk material with a smooth surface.
 
\begin{figure}
\centering
   \includegraphics[width=\hsize]{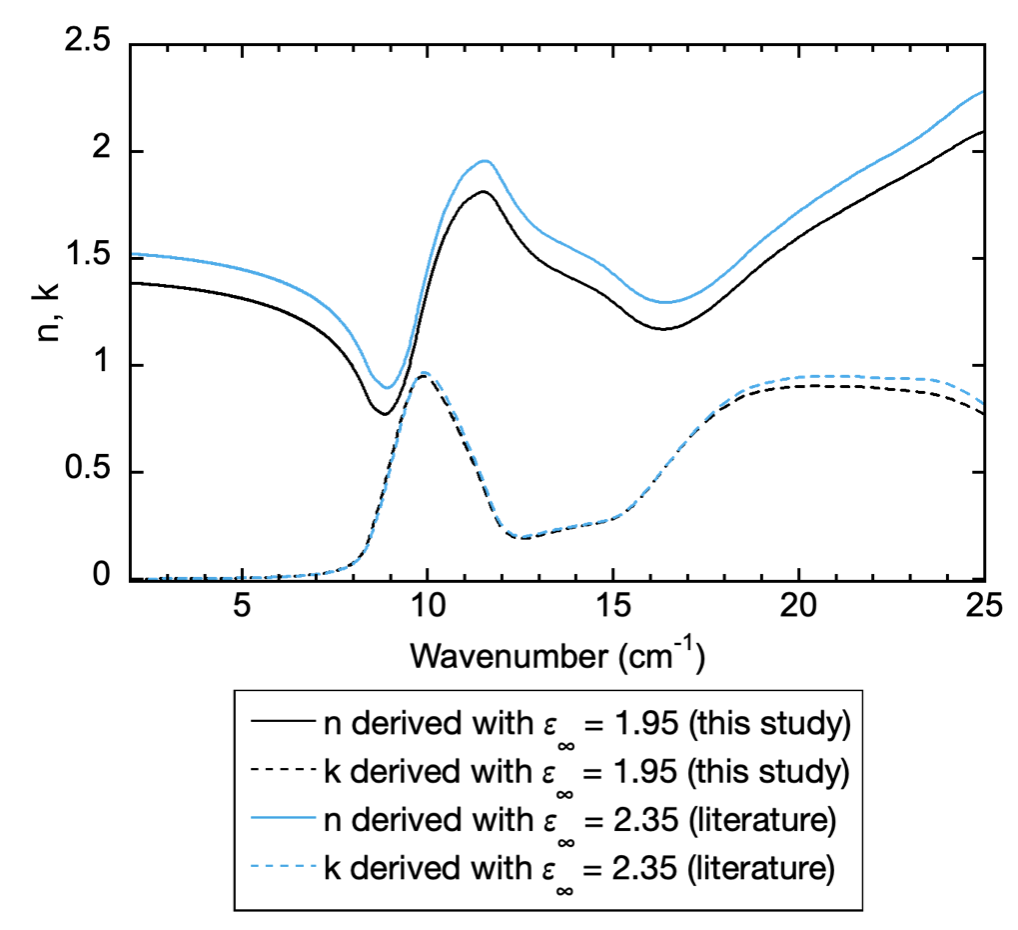}
     \caption{Refractive index ($n$; upper panel) and the extinction coefficient ($k$; bottom panel) derived using the $\epsilon_\infty= 1.95$ determined from reflectance measured in this study and the $\epsilon_\infty=2.35$ determined from reflectance measured in previous studies \citep{Jaeger2003, Dorschner1995}.}
     \label{fig:nk_eps}
\end{figure}

\begin{figure}
\centering
   \includegraphics[width=\hsize]{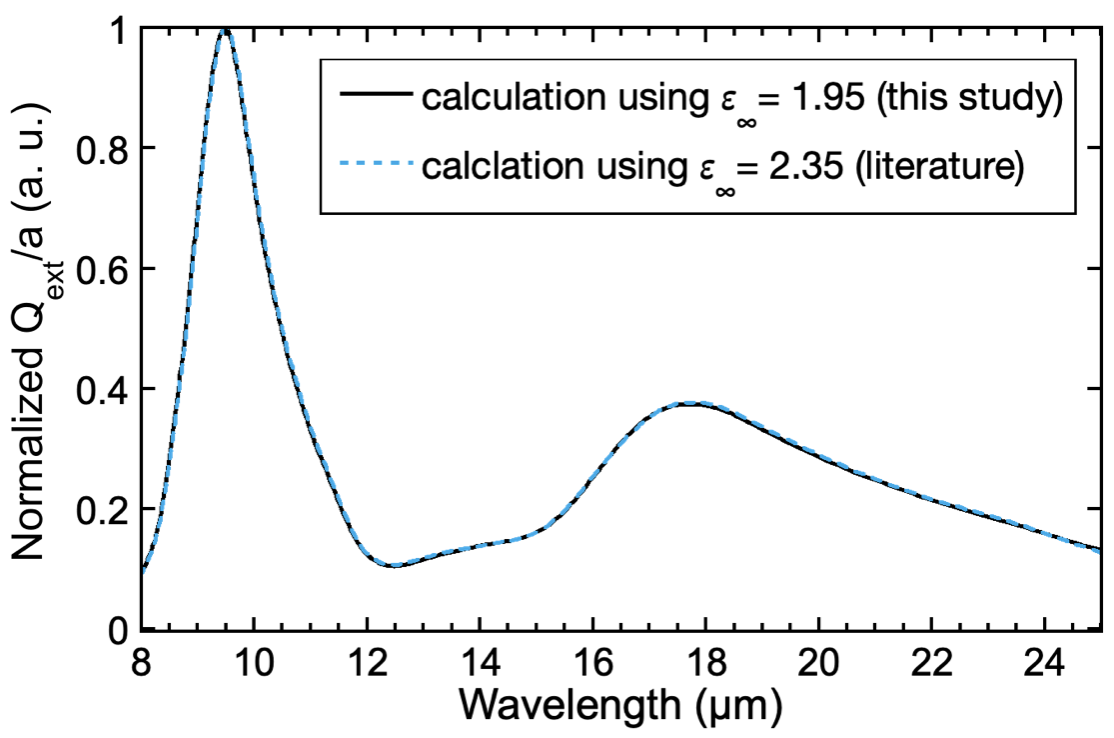}
     \caption{Normalized spectra of $Q_{\mathrm{ext}}/a$ in vacuum calculated using the $\epsilon_\infty = 1.95$ determined from reflectance measured in this study (black line) and the $\epsilon_\infty = 2.35$ determined from reflectance measured in previous studies \citep[blue line;][]{Jaeger2003, Dorschner1995}.}
     \label{fig:Qa_eps}
\end{figure}

 To estimate the effects of different reflectance at high wavenumbers on the determination of $\epsilon_\infty$ and optical constants, we also derived the optical constants for CI-2 (Ca$_{0.05}$Mg$_{0.92}$Na$_{0.06}$Al$_{0.07}$Si) by adopting $\epsilon_\infty = 2.35$, a value consistent with the high-wavenumber reflectance reported for MgSiO$_3$ glass \citep{Jaeger2003, Dorschner1995}.
 A comparison of derived optical constants and normalized spectra of $Q_{\mathrm{ext}}/a$ in vacuum using $\epsilon_\infty= 1.95$ and $\epsilon_\infty=2.35$ is shown in Fig. \ref{fig:nk_eps} and Fig. \ref{fig:Qa_eps}. The refractive index $n$ derived with $\epsilon_\infty =1.95$ was slightly lower than that determined with $\epsilon_\infty =2.35$, while the extinction coefficient $k$ showed little difference. In normalized spectra of $Q_{\mathrm{ext}}/a$, the 10 and 18~$\mu$m peak exhibited negligible shifts, within $0.1 \mathrm{\, \mu m}$ (from 9.42 to $9.45 \mathrm{\, \mu m}$ and from 17.8 to $17.9 \mathrm{\, \mu m}$) when $\epsilon_\infty$ varied from 1.95 to 2.35. These results indicate that a $\sim$36$\%$  difference in reflectance at high wavenumbers has a limited impact on spectral changes. However, further studies are necessary to refine methods for measuring the reflectance of nanoparticles and determining optical constants with higher accuracy.

\subsection{Comparison of measured and modeled spectra of amorphous silicate particles with metallic iron}
 The STEM analysis showed that the Fe-Ni-including product with CI chondritic composition (CI-1) had a metallic core and silicate shell structure with the averaged core size ratio ($r_\mathrm{c}/r$) of $\sim0.50$ (Figs. \ref{fig:TEM} and \ref{fig:rc}). In contrast, the product in run CI-2 was pure amorphous silicate particles. The amorphous silicate shell of CI-1 and the silicate grain of CI-2 had almost the same chemical composition (Table \ref{table:STEM}). Despite the structural difference between CI-1 and CI-2, the peak position of the amorphous silicate feature did not change as a result of the absorption measurement (Fig. \ref{fig:Abs}). 
 
\begin{figure}
\centering
   \includegraphics[width=\hsize]{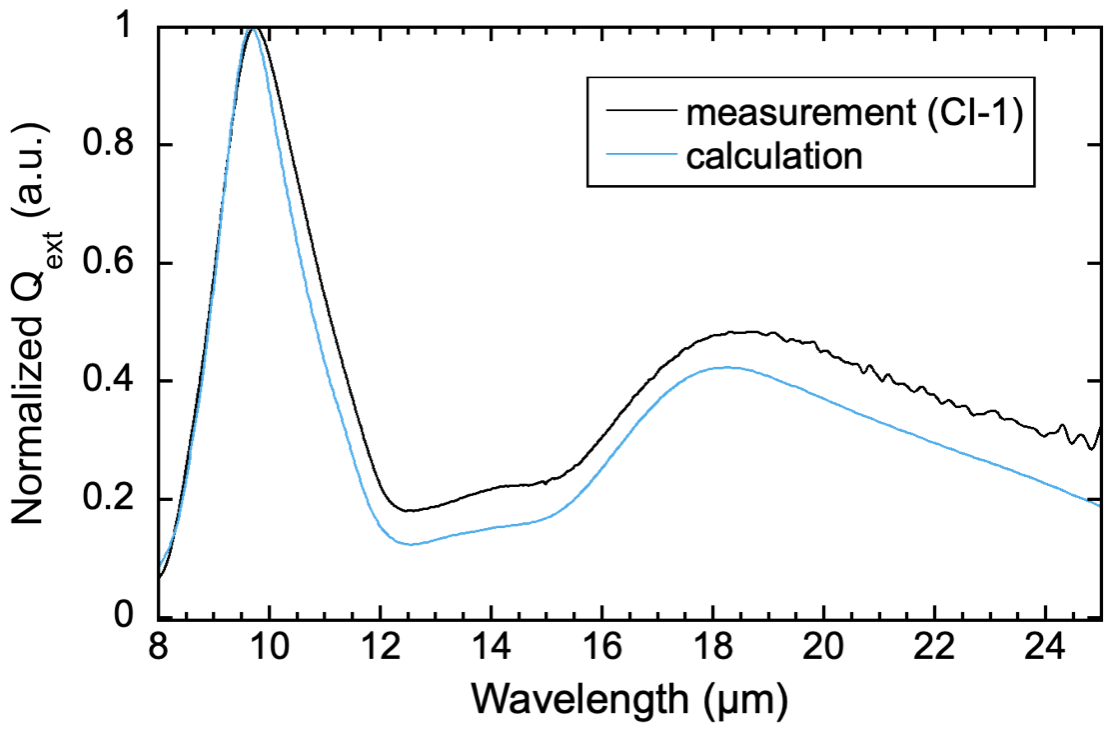}
     \caption{Measured and calculated spectra of amorphous silicate particles with metallic iron cores. The black line shows the measurement of the product with a kamacite core (CI-1), and the blue line shows the calculated spectrum assuming amorphous silicate with a metallic core ($r_\mathrm{c}/r = 0.50$).}
     \label{fig:Qa_meas_model}
\end{figure}

 We calculated the extinction coefficient of amorphous silicates with metallic Fe-Ni cores using the optical constants of the iron-free product with the CI chondritic composition (CI-2) obtained in Sect. 3.4. and $\alpha$-iron \citep{Ordal1985}. Here, we assumed $r_\mathrm{c}/r = 0.50$ and the medium effect of KBr to compare with the measured absorption spectrum of CI-1 nanoparticles dispersed in a KBr pellet. In the calculation, we used the FORTRAN code DMiLay \citep{Ackerman1981}. This code computes electromagnetic scattering by a particle composed of a spherical core and a spherical shell without assumptions such as effective medium approximations. The measured spectrum of the product in the run CI-1 and the calculated spectrum of particles with metallic core ($r_\mathrm{c}/r = 0.50$) were compared in Fig. \ref{fig:Qa_meas_model}. The optical constants of metallic iron were taken from the literature \citep{Ordal1985}. There was a difference of about $\sim 0.1  \mathrm{\, \mu m}$ in the peak position between the measured and modeled spectra.
 
\begin{figure}
\centering
   \includegraphics[width=\hsize]{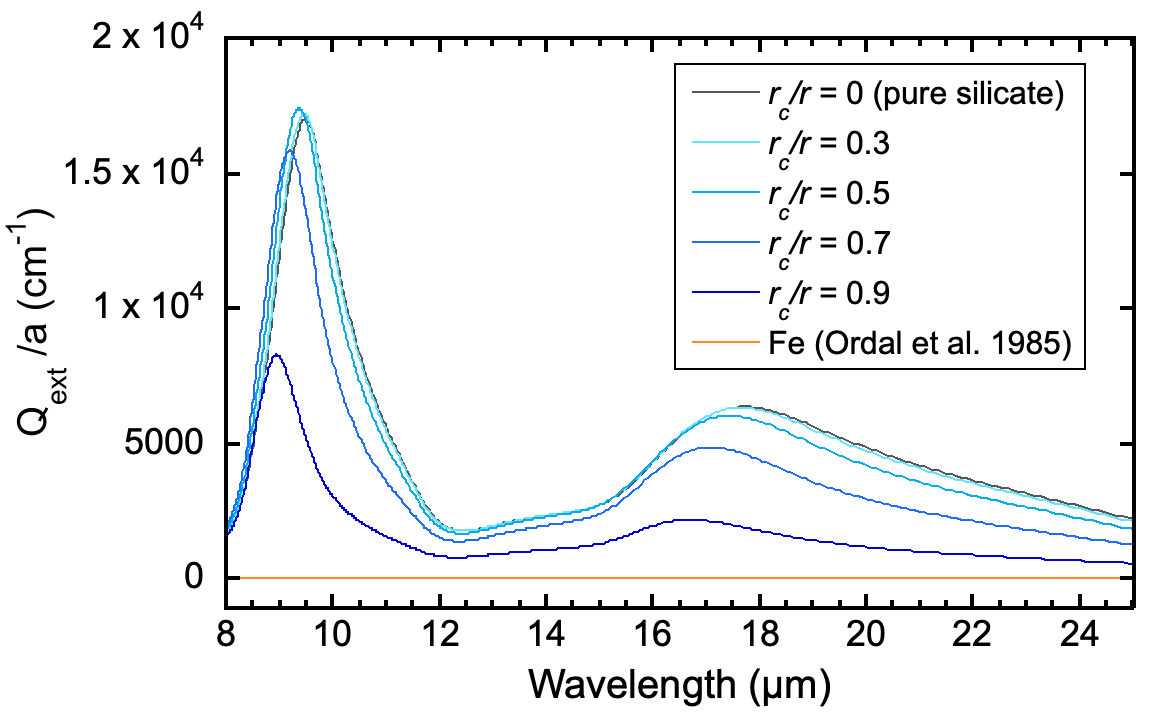}
     \caption{Spectra of amorphous silicate dust with a metallic iron core and various core–shell ratios calculated using DMiLay \citep[gray and blue lines;][]{Ackerman1981}. The orange line shows the $Q_{\mathrm{ext}}/a$ of iron calculated with optical constants taken from the literature \citep{Ordal1985}.}
     \label{fig:DMilay_Fe}
\end{figure}

 This discrepancy is likely due to the assumption in the calculation that all dust grains have an identical value of $r_\mathrm{c}/r = 0.50$, whereas the experimental products exhibit a distribution of core size ratio, as shown in Fig. \ref{fig:rc}. It should also be noted that there is a size-dependent bias in the STEM observations. In STEM-EDS analysis, only particles larger than $\sim\phi 50 \mathrm{\, nm}$ were counted because X-ray counts obtained from smaller particles without electron beam damage were insufficient to determine the boundaries between the metallic core and silicate mantle. In addition, many small particles ($< \phi50 \mathrm{\, nm}$) were stuck to the larger grains, and their metallic cores were not confirmed in the STEM dark field images (Fig. \ref{fig:TEM}). Figure \ref{fig:DMilay_Fe} shows the calculated $Q_{\mathrm{ext}}/a$ spectra of core-shell dust with various radius ratios. The peak positions do not change dramatically for grains with $r_\mathrm{c}/r < 0.3$, whereas the peak positions of the grains with $r_\mathrm{c}/r > 0.3$ significantly shift to the shorter wavelength. The identical MIR features observed in CI-1 and CI-2, despite the presence of metallic cores in some large silicate grains, suggest that the IR absorption spectrum is predominantly governed by numerous small particles ($<\phi50 \mathrm{\, nm}$), for which metallic cores were not confirmed.
 
 The elemental abundance of iron in our product CI-1 (Fe/Si of $\sim$0.82) was almost the CI chondritic composition (Fe/Si of $\sim$0.85). Contrary to the intuitive expectation that the metallic iron of the solar abundance would have a significant impact on the spectra, our results suggest that the spectral feature is predominantly controlled by the number density and size distribution of metallic inclusions rather than their total elemental abundance. 

 Some chondritic porous interplanetary dust particles show IR spectral features similar to those of circumstellar and interstellar amorphous silicate dust (Bradley et al. 1999). Although the observed GEMS are aggregated and their grain boundaries are hard to distinguish, the size of metallic iron particles distributed from approximately 5 to $25 \mathrm{\, nm}$ in the silicate aggregates ($\sim 50 \mathrm{\, nm}$; Fig. 3 of \citealt{Bradley1999}). The few observed isolated GEMS particles showed core-shell ratios $r_\mathrm{c}/r$ varying from 0.35 to 0.50 (Fig. 8 of \citealt{Messenger2015}). Like the experimental product CI-1, metallic iron particles embedded in GEMS exhibit a size distribution. As suggested by our experimental results, metallic particles with a size distribution are difficult to detect in the MIR spectra when observed in space. Therefore, circumstellar dust exhibiting MIR spectral features similar to those of GEMS may also contain metallic particles. As the presence of metallic cores enhances absorption efficiency in NIR wavelengths \citep{Jaeger2003}, optical constants of amorphous silicate with a metallic iron core need to be determined at a wider wavelength to further verify the presence of metallic cores in circumstellar dust.
 
Mg-silicate condenses at a higher temperature than metallic iron because of its low surface energy under rapid cooling conditions\citep{Yamamoto1977}. The spherical shape of the condensed grains indicates that amorphous silicate condensed as melt droplets, as proposed in previous studies \citep{Kim2021, Matsuno2021, Enju2022}. In the run CI-1, metallic iron particles condense on the surface and then move into the inside of the silicate melt \citep{Matsuno2021, Kim2021}. The iron particles in the silicate melt coagulate to grow into a large core to minimize their surface energy, leading to a heterogeneous size distribution from core-less amorphous silicate grains to large cores ($r_\mathrm{c}/r \sim 0.87$) in relatively large silicate grains as seen in our experiment. This can reduce the interaction cross section of the metal particles with light and diminish their spectral contribution. In this way, the presence of metallic iron may be "hidden" in the IR spectra despite its high elemental abundance. Such processes could occur in rapidly cooled dense gas, like the rear of shock waves in circumstellar environments or chondrule-forming regions in the protosolar disk.

\section{Conclusions}
 We performed condensation experiments using the ITP system in the Mg–Ca–Na–Al–Si–Fe–Ni–O and Mg–Ca–Na–Al–Si–O systems to synthesize two types of amorphous silicate dust analogs with a solar composition: amorphous silicate with a metallic Fe-Ni core, which has a core size proportion ($r_\mathrm{c}/r$) from 0 to 0.87, and Fe-Ni-free amorphous silicate. Reflectance spectra were measured in addition to absorbance spectra to derive the optical constants. The powders were pressed into pellets, and the reference materials were prepared by coating the surface of the sample pellets with gold. This approach effectively prevented the decrease in reflectance that would have been caused by the surface roughness. We derived optical constants by determining $\epsilon_\infty$ from the reflectance at $> 4000$ cm$^{-1}$ and fitting the $Q_{\mathrm{ext}}/a$ calculated from the model to that derived from the measured absorbance.

 Using the determined optical constants, the effects of metallic iron particles in amorphous silicate grains on IR spectral features were discussed. The modeled spectra of core-shell dust showed clear shifts of the Si--O stretching and Si--O--Si bending peaks with larger metallic cores ($r_\mathrm{c}/r > 0.3$). However, the peak positions  of the absorbance spectra of the experimental products remained the same regardless of the presence or absence of metallic cores. This is because Fe and Ni were concentrated in metallic cores ($r_\mathrm{c}/r > 0.3$) in large amorphous silicate grains ($> \phi 50 \mathrm{\, nm}$) and because small amorphous silicate grains ($< \phi 50 \mathrm{\, nm}$) with small cores ($r_\mathrm{c}/r > 0.3$) or without cores dominate the spectral features. As a result, the presence of metallic iron particles cannot be detected from IR spectra. The spectral effects of metallic iron depend more on the size distribution and number density of inclusions than on their elemental abundance. Metallic Fe-Ni grains formed on the surface of silicate melts sink into the silicate melts and coagulate into a large core to minimize surface energy, thereby leading to a size distribution of metallic cores.
 
 The metallic-iron-bearing dust analog produced in this study is comparable to GEMS in terms of iron abundance and the distribution of metallic core sizes. However, even in the presence of abundant iron, the MIR spectra are indistinguishable from those of pure amorphous silicate dust when the metallic cores have a size distribution. This implies that the dust around AGB stars is not composed solely of pure amorphous silicates but could also contain metallic iron cores similar to those in GEMS. Combining radiative transfer modeling using the optical constants obtained in this study with MIR and NIR spectroscopic observations provides a powerful means to constrain the presence of circumstellar GEMS-like materials.

\begin{acknowledgements}
      H. Enomoto and A. Takigawa were financially supported by JSPS KAKENHI Grant Number 23H01213. H. Enomoto was supported by JST SPRING, Grant Number JPMJSP2108. The authors thank Dr. Ryo Tazaki (The University of Tokyo) for constructive discussions on spectral modeling of dust with metallic cores. We also thank an anonymous referee for their insightful comments that improved the quality of this paper.
\end{acknowledgements}

\bibliographystyle{aa}
\bibliography{references}

@article{Ackerman1981,
  author = {Ackerman, T. P. and Toon, O. B.},
  year = {1981},
  title = {Absorption of visible radiation in atmosphere containing mixtures of absorbing and nonabsorbing particles},
  journal = {Applied Optics},
  volume = {20},
  number = {20},
  pages = {3661--3668}
}

@book{Bohren1983,
  author = {Bohren, C. F. and Huffman, D. R.},
  year = {1983},
  title = {Absorption and Scattering of Light by Small Particles},
  publisher = {Wiley},
  address = {New York}
}

@article{Bradley1999,
  author = {Bradley, J. P. and Keller, L. P. and Snow, T. P. and Hanner, M. S. and Flynn, G. J. and Gezo, J. C. and Bowey, J. E.},
  year = {1999},
  title = {An infrared spectral match between GEMS and interstellar grains},
  journal = {Science},
  volume = {285},
  number = {5434},
  pages = {1716--1718}
}

@article{Cabedo2024,
  author = {Cabedo, V. and Pareras, G. and Allitt, J. and Rimola, A. and Llorca, J. and Yiu, H. H. P. and McCoustra, M. R.},
  year = {2024},
  title = {Reactivity of chondritic meteorites under H2-rich atmospheres: formation of H2S},
  journal = {Monthly Notices of the Royal Astronomical Society},
  volume = {535},
  number = {3},
  pages = {2714--2723}
}

@article{Dorschner1995,
  author = {Dorschner, J. and Begemann, B. and Henning, T. and Jaeger, C. and Mutschke, H.},
  year = {1995},
  title = {Steps toward interstellar silicate mineralogy. II. Study of Mg-Fe-silicate glasses of variable composition},
  journal = {Astronomy and Astrophysics},
  volume = {300},
  pages = {503}
}

@article{Draine1984,
  author = {Draine, B. T. and Lee, H. M.},
  year = {1984},
  title = {Optical properties of interstellar graphite and silicate grains},
  journal = {Astrophysical Journal},
  volume = {285},
  pages = {89}
}

@article{Enju2022,
  author = {Enju, S. and Kawano, H. and Tsuchiyama, A. and Kim, T. H. and Takigawa, A. and Matsuno, J. and Komaki, H.},
  year = {2022},
  title = {Condensation of cometary silicate dust using an induction thermal plasma system. II. Mg-Fe-Si-O-S system and the effects of sulfur and redox conditions},
  journal = {Astronomy \& Astrophysics},
  volume = {661},
  pages = {A121}
}

@article{Fabian2001,
  author = {Fabian, D. and Henning, T. and Jäger, C. and Mutschke, H. and Dorschner, J. and Wehrhan, O.},
  year = {2001},
  title = {Steps toward interstellar silicate mineralogy-VI. Dependence of crystalline olivine IR spectra on iron content and particle shape},
  journal = {Astronomy \& Astrophysics},
  volume = {378},
  number = {1},
  pages = {228--238}
}

@article{Gail1999,
  author = {Gail, H. P. and Sedlmayr, E.},
  year = {1999},
  title = {Mineral formation in stellar winds. I. Condensation sequence of silicate and iron grains in stationary oxygen rich outflows},
  journal = {Astronomy \& Astrophysics},
  volume = {347},
  pages = {594--616}
}

@article{Henning2010,
  author = {Henning, T.},
  year = {2010},
  title = {Cosmic Silicates},
  journal = {Annual Review of Astronomy and Astrophysics},
  volume = {48},
  pages = {21--46}
}

@article{Hofner2018,
  author  = {Höfner, Susanne and Olofsson, Hans},
  title   = {Mass loss of stars on the asymptotic giant branch: Mechanisms, models and measurements},
  journal = {The Astronomy and Astrophysics Review},
  year    = {2018},
  volume  = {26},
  number  = {1},
  pages   = {1},
  doi     = {10.1007/s00159-017-0106-5}
}

@article{Jaeger2003,
  author = {Jäger, C. and Dorschner, J. and Mutschke, H. and Posch, T. and Henning, T.},
  year = {2003},
  title = {Steps toward interstellar silicate mineralogy-VII. Spectral properties and crystallization behaviour of magnesium silicates produced by the sol-gel method},
  journal = {Astronomy \& Astrophysics},
  volume = {408},
  number = {1},
  pages = {193--204}
}

@article{Jaeger1994,
  author = {Jäger, C. and Mutschke, H. and Begemann, B. and Dorschner, J. and Henning, T.},
  year = {1994},
  title = {Steps toward interstellar silicate mineralogy. I: Laboratory results of a silicate glass of mean cosmic composition},
  journal = {Astronomy \& Astrophysics},
  volume = {292},
  number = {2},
  pages = {641--655}
}

@article{Jones1976,
  author = {Jones, T. W. and Merrill, K. M.},
  year = {1976},
  title = {Model dust envelopes around late-type stars},
  journal = {Astrophysical Journal},
  volume = {209},
  pages = {509--524}
}

@article{Keller2011,
  author = {Keller, L. P. and Messenger, S.},
  year = {2011},
  title = {On the origins of GEMS grains},
  journal = {Geochimica et Cosmochimica Acta},
  volume = {75},
  number = {18},
  pages = {5336--5365}
}

@article{Kemper2002,
  author = {Kemper, F. and De Koter, A. and Waters, L. B. F. M. and Bouwman, J. and Tielens, A. G. G. M.},
  year = {2002},
  title = {Dust and the spectral energy distribution of the OH/IR star OH 127.8+ 0.0: Evidence for circumstellar metallic iron},
  journal = {Astronomy \& Astrophysics},
  volume = {384},
  number = {2},
  pages = {585--593}
}

@article{Kemper2004,
  author = {Kemper, F. and Vriend, W. J. and Tielens, A. G. G. M.},
  year = {2004},
  title = {The absence of crystalline silicates in the diffuse interstellar medium},
  journal = {The Astrophysical Journal},
  volume = {609},
  number = {2},
  pages = {826}
}

@article{Kemper2005,
  author = {Kemper, F. and Vriend, W. J. and Tielens, A. G. G. M.},
  year = {2005},
  title = {ERRATUM: ‘‘THE ABSENCE OF CRYSTALLINE SILICATES IN THE DIFFUSE INTERSTELLAR MEDIUM’’ (ApJ, 609, 826 [2004])},
  journal = {The Astrophysical Journal},
  volume = {633},
  pages = {534}
}

@article{Kim2021,
  author = {Kim, T. H. and Takigawa, A. and Tsuchiyama, A. and Matsuno, J. and Enju, S. and Kawano, H. and Komaki, H.},
  year = {2021},
  title = {Condensation of cometary silicate dust using an induction thermal plasma system: I. Enstatite and CI chondritic composition},
  journal = {Astronomy \& Astrophysics},
  volume = {656},
  pages = {A42}
}

@article{Koike1989,
  author = {Koike, C. and Hasegawa, H. and Asada, N. and Komatuzaki, T.},
  year = {1989},
  title = {Optical constants of fine particles for the infrared region},
  journal = {Monthly Notices of the Royal Astronomical Society},
  volume = {239},
  number = {1},
  pages = {127--137}
}

@article{Koike1995,
  author = {Koike, C. and Kaito, C. and Yamamoto, T. and Shibai, H. and Kimura, S. and Suto, H.},
  year = {1995},
  title = {Extinction spectra of corundum in the wavelengths from UV to FIR},
  journal = {Icarus},
  volume = {114},
  number = {1},
  pages = {203--214}
}

@article{Kozasa1987,
  title={Grain formation through nucleation process in astrophysical environments. II: Nucleation and grain growth accompanied by chemical reaction},
  author={Kozasa, Takahashi and Hasegawa, Hiroichi},
  journal={Progress of Theoretical Physics},
  volume={77},
  number={6},
  pages={1402--1410},
  year={1987},
  publisher={Oxford University Press}
}

@incollection{Lodders1999,
  author = {Lodders, K. and Fegley Jr, B.},
  year = {1999},
  title = {Condensation chemistry of circumstellar grains},
  booktitle = {Symposium-International Astronomical Union},
  volume = {191},
  pages = {279--290},
  publisher = {Cambridge University Press}
}

@book{Lodders2009,
  author = {Lodders, K. and Palme, H. and Gail, H. P.},
  year = {2009},
  title = {4.4 Abundances of the elements in the Solar System: 4 The Solar System},
  journal = {Solar system},
  pages = {712--770}
}

@article{Matsuno2022,
  author = {Matsuno, J. and Tsuchiyama, A. and Miyake, A. and Nakamura-Messenger, K. and Messenger, S.},
  year = {2022},
  title = {Three-dimensional observation of GEMS grains: Their high-temperature condensation origin},
  journal = {Geochimica et Cosmochimica Acta},
  volume = {320},
  pages = {207--222}
}

@article{Matsuno2021,
  author = {Matsuno, J. and Tsuchiyama, A. and Watanabe, T. and Tanaka, M. and Takigawa, A. and Enju, S. and Koike, C. and Chihara, H. and Miyake, A.},
  year = {2021},
  title = {Condensation of glass with multimetal nanoparticles: Implications to a formation process of GEMS grains},
  journal = {Astrophysical Journal},
  volume = {911},
  pages = {47--62}
}

@article{Messenger2015,
  author = {Messenger, S. and Nakamura-Messenger, K. and Keller, L. P. and Clemett, S. J.},
  year = {2015},
  title = {Pristine stratospheric collection of interplanetary dust on an oil-free polyurethane foam substrate},
  journal = {Meteoritics \& Planetary Science},
  volume = {50},
  number = {8},
  pages = {1468--1485}
}

@article{Min2006,
  author = {Min, M. and Dominik, C. and Hovenier, J. W. and de Koter, A. and Waters, L. B. F. M.},
  year = {2006},
  title = {The 10 $\mu$m amorphous silicate feature of fractal aggregates and compact particles with complex shapes},
  journal = {Astronomy \& Astrophysics},
  volume = {445},
  number = {3},
  pages = {1005--1014}
}

@article{Min2005,
  author = {Min, M. and Hovenier, J. W. and de Koter, A.},
  year = {2005},
  title = {Modeling optical properties of cosmic dust grains using a distribution of hollow spheres},
  journal = {Astronomy \& Astrophysics},
  volume = {432},
  number = {3},
  pages = {909--920}
}

@article{Mutschke1998,
  author = {Mutschke, H. and Begemann, B. and Dorschner, J. and Guertler, J. and Gustafson, B. and Henning, T. and Stognienko, R.},
  year = {1998},
  title = {Steps toward interstellar silicate mineralogy. III. The role of aluminium in circumstellar amorphous silicates},
  journal = {Astronomy and Astrophysics},
  volume = {333},
  pages = {188--198}
}

@article{Nuth2002,
  author = {Nuth, J. A. and Rietmeijer, F. J. and Hill, H. G.},
  year = {2002},
  title = {Condensation processes in astrophysical environments: The composition and structure of cometary grains},
  journal = {Meteoritics \& Planetary Science},
  volume = {37},
  number = {11},
  pages = {1579--1590}
}

@article{Nuth1990,
  author = {Nuth, J. A. and Hecht, J. H.},
  year = {1990},
  title = {Signatures of aging silicate dust},
  journal = {Astrophysics and Space Science},
  volume = {163},
  pages = {79--94}
}

@article{Ordal1985,
  author = {Ordal, M. A. and Bell, R. J. and Alexander, R. W. and Long, L. L. and Querry, M. R.},
  year = {1985},
  title = {Optical properties of fourteen metals in the infrared and far infrared: Al, Co, Cu, Au, Fe, Pb, Mo, Ni, Pd, Pt, Ag, Ti, V, and W},
  journal = {Applied Optics},
  volume = {24},
  number = {24},
  pages = {4493--4499}
}

@article{Ossenkopf1992,
  author = {Ossenkopf, V. and Henning, T. and Mathis, J. S.},
  year = {1992},
  title = {Constraints on cosmic silicates},
  journal = {Astronomy and Astrophysics},
  volume = {261},
  pages = {567--578}
}

@article{Reboussin2014,
  author = {Reboussin, L. and Wakelam, V. and Guilloteau, S. and Hersant, F.},
  year = {2014},
  title = {Grain-surface reactions in molecular clouds: the effect of cosmic rays and quantum tunnelling},
  journal = {Monthly Notices of the Royal Astronomical Society},
  volume = {440},
  number = {4},
  pages = {3557--3567}
}

@article{Rietmeijer2002,
  author = {Rietmeijer, F. J. and Hallenbeck, S. L. and Nuth III, J. A. and Karner, J. M.},
  year = {2002},
  title = {Amorphous magnesiosilicate smokes annealed in vacuum: The evolution of magnesium silicates in circumstellar and cometary dust},
  journal = {Icarus},
  volume = {156},
  number = {1},
  pages = {269--286}
}

@article{Rogers1983,
  author = {Rogers, C. and Martin, P. G. and Crabtree, D. R.},
  year = {1983},
  title = {The circumstellar dust of MU Cephei},
  journal = {Astrophysical Journal, Part 1},
  volume = {272},
  number = {1},
  pages = {175--181}
}

@article{Speck2015,
  author = {Speck, A. K. and Pitman, K. M. and Hofmeister, A. M.},
  year = {2015},
  title = {Better alternatives to ``astronomical silicate'': Laboratory-based optical functions of chondritic/solar abundance glass with application to HD 161796},
  journal = {The Astrophysical Journal},
  volume = {809},
  number = {1},
  pages = {65}
}

@article{Speck2011,
  author = {Speck, A. K. and Whittington, A. G. and Hofmeister, A. M.},
  year = {2011},
  title = {Disordered silicates in space: a study of laboratory spectra of ``amorphous'' silicates},
  journal = {The Astrophysical Journal},
  volume = {740},
  number = {2},
  pages = {93}
}

@article{Takigawa2019,
  author = {Takigawa, A. and Kim, T. H. and Igami, Y. and Umemoto, T. and Tsuchiyama, A. and Koike, C. and Matsuno, J. and Watanabe, T.},
  year = {2019},
  title = {Formation of Transition Alumina Dust around Asymptotic Giant Branch Stars: Condensation Experiments Using Induction Thermal Plasma Systems},
  journal = {Astrophysical Journal Letters},
  volume = {878},
  number = {1},
  pages = {1--8}
}

@article{Tamanai2006,
  author = {Tamanai, A. and Mutschke, H. and Blum, J. and Meeus, G.},
  year = {2006},
  title = {The 10 $\mu$m infrared band of silicate dust: A laboratory study comparing the aerosol and KBr pellet techniques},
  journal = {The Astrophysical Journal},
  volume = {648},
  number = {2},
  pages = {L147}
}

@book{Tielens2005,
  author = {Tielens, A. G. G. M.},
  year = {2005},
  title = {The physics and chemistry of the interstellar medium},
  publisher = {Cambridge University Press}
}

@inproceedings{Tielens2018,
  author = {Tielens, A. G. G. M.},
  year = {2018},
  title = {The Cycling of Matter between Stars, the Interstellar Medium, and the Intergalactic Medium},
  booktitle = {JAXA Special Publication: Proceedings of the SPICA Science Conference from Exoplanets to Distant Galaxies: SPICA's New Window on the Cool Universe},
  volume = {17},
  pages = {163--171}
}

@article{Woitke2006,
  title={Too little radiation pressure on dust in the winds of oxygen-rich AGB stars},
  author={Woitke, Peter},
  journal={Astronomy \& Astrophysics},
  volume={460},
  number={2},
  pages={L9--L12},
  year={2006},
  publisher={EDP Sciences}
}

@article{Yamamoto1977,
  author = {Yamamoto, T. and Hasegawa, H.},
  year = {1977},
  title = {Grain formation through nucleation process in astrophysical environment},
  journal = {Progress of Theoretical Physics},
  volume = {58},
  number = {3},
  pages = {816--828}
}

@article{Zeidler2013,
  author = {Zeidler, S. and Posch, T. and Mutschke, H.},
  year = {2013},
  title = {Optical constants of refractory oxides at high temperatures-Mid-infrared properties of corundum, spinel, and $\alpha$-quartz, potential carriers of the 13 $\mu$m feature},
  journal = {Astronomy \& Astrophysics},
  volume = {553},
  pages = {A81}
}

@article{Zhukovska2018,
  author = {Zhukovska, S. and Henning, T. and Dobbs, C.},
  year = {2018},
  title = {Iron and silicate dust growth in the galactic interstellar medium: clues from element depletions},
  journal = {The Astrophysical Journal},
  volume = {857},
  number = {2},
  pages = {94}
}

\begin{appendix} 
\onecolumn
\section{Optical constants of the metallic iron-free product CI-2 (Ca$_{0.05}$Mg$_{0.92}$Na$_{0.06}$Al$_{0.07}$Si)} 

\begin{longtable}{c c c c c c c c c c c }
\caption{Optical constants of sample CI-2 (Ca$_{0.05}$Mg$_{0.92}$Na$_{0.06}$Al$_{0.07}$Si).}\\
\label{table:nk}

        Wavelength ($\mu$m) & n & k & ~ & Wavelength ($\mu$m) & n & k & ~ & Wavelength ($\mu$m) & n & k \\ 
\hline
\endfirsthead
\caption{Continued.}\\
\hline
        Wavelength ($\mu$m) & n & k & ~ & Wavelength ($\mu$m) & n & k & ~ & Wavelength ($\mu$m) & n & k \\ 
\hline
\endhead
\hline
\endfoot
        25.05 & 2.099 & 0.764 & ~ & 19.57 & 1.546 & 0.894 & ~ & 16.05 & 1.176 & 0.446 \\ 
        24.93 & 2.091 & 0.776 & ~ & 19.49 & 1.536 & 0.893 & ~ & 16.00 & 1.178 & 0.435 \\ 
        24.81 & 2.082 & 0.787 & ~ & 19.42 & 1.527 & 0.891 & ~ & 15.96 & 1.181 & 0.425 \\ 
        24.69 & 2.072 & 0.798 & ~ & 19.35 & 1.518 & 0.889 & ~ & 15.91 & 1.185 & 0.415 \\ 
        24.58 & 2.062 & 0.808 & ~ & 19.28 & 1.508 & 0.887 & ~ & 15.86 & 1.188 & 0.405 \\ 
        24.46 & 2.051 & 0.817 & ~ & 19.21 & 1.499 & 0.885 & ~ & 15.81 & 1.192 & 0.395 \\ 
        24.35 & 2.040 & 0.826 & ~ & 19.13 & 1.489 & 0.883 & ~ & 15.76 & 1.196 & 0.386 \\ 
        24.23 & 2.028 & 0.833 & ~ & 19.06 & 1.480 & 0.881 & ~ & 15.71 & 1.201 & 0.376 \\ 
        24.12 & 2.016 & 0.840 & ~ & 18.99 & 1.470 & 0.879 & ~ & 15.67 & 1.206 & 0.368 \\ 
        24.01 & 2.004 & 0.847 & ~ & 18.93 & 1.460 & 0.876 & ~ & 15.62 & 1.211 & 0.359 \\ 
        23.90 & 1.992 & 0.852 & ~ & 18.86 & 1.450 & 0.873 & ~ & 15.57 & 1.216 & 0.351 \\ 
        23.79 & 1.981 & 0.857 & ~ & 18.79 & 1.440 & 0.870 & ~ & 15.53 & 1.222 & 0.343 \\ 
        23.68 & 1.969 & 0.862 & ~ & 18.72 & 1.429 & 0.867 & ~ & 15.48 & 1.228 & 0.336 \\ 
        23.57 & 1.957 & 0.865 & ~ & 18.65 & 1.419 & 0.863 & ~ & 15.43 & 1.234 & 0.329 \\ 
        23.46 & 1.946 & 0.868 & ~ & 18.59 & 1.409 & 0.859 & ~ & 15.39 & 1.240 & 0.322 \\ 
        23.36 & 1.935 & 0.871 & ~ & 18.52 & 1.398 & 0.854 & ~ & 15.34 & 1.246 & 0.316 \\ 
        23.25 & 1.925 & 0.874 & ~ & 18.45 & 1.388 & 0.849 & ~ & 15.30 & 1.253 & 0.310 \\ 
        23.15 & 1.914 & 0.876 & ~ & 18.39 & 1.378 & 0.844 & ~ & 15.25 & 1.259 & 0.305 \\ 
        23.05 & 1.904 & 0.878 & ~ & 18.32 & 1.368 & 0.838 & ~ & 15.21 & 1.265 & 0.300 \\ 
        22.94 & 1.894 & 0.880 & ~ & 18.26 & 1.358 & 0.833 & ~ & 15.16 & 1.272 & 0.295 \\ 
        22.84 & 1.885 & 0.882 & ~ & 18.19 & 1.348 & 0.826 & ~ & 15.12 & 1.278 & 0.291 \\ 
        22.74 & 1.875 & 0.883 & ~ & 18.13 & 1.338 & 0.820 & ~ & 15.07 & 1.284 & 0.287 \\ 
        22.64 & 1.866 & 0.885 & ~ & 18.07 & 1.329 & 0.813 & ~ & 15.03 & 1.290 & 0.284 \\ 
        22.55 & 1.857 & 0.886 & ~ & 18.01 & 1.320 & 0.806 & ~ & 14.99 & 1.297 & 0.280 \\ 
        22.45 & 1.848 & 0.888 & ~ & 17.94 & 1.311 & 0.798 & ~ & 14.94 & 1.303 & 0.278 \\ 
        22.35 & 1.839 & 0.890 & ~ & 17.88 & 1.302 & 0.791 & ~ & 14.90 & 1.308 & 0.275 \\ 
        22.26 & 1.830 & 0.891 & ~ & 17.82 & 1.293 & 0.783 & ~ & 14.86 & 1.314 & 0.273 \\ 
        22.16 & 1.821 & 0.893 & ~ & 17.76 & 1.285 & 0.775 & ~ & 14.82 & 1.319 & 0.270 \\ 
        22.07 & 1.812 & 0.894 & ~ & 17.70 & 1.276 & 0.766 & ~ & 14.77 & 1.325 & 0.268 \\ 
        21.97 & 1.803 & 0.895 & ~ & 17.64 & 1.268 & 0.758 & ~ & 14.73 & 1.330 & 0.267 \\ 
        21.88 & 1.794 & 0.897 & ~ & 17.58 & 1.261 & 0.749 & ~ & 14.69 & 1.335 & 0.265 \\ 
        21.79 & 1.785 & 0.898 & ~ & 17.52 & 1.253 & 0.740 & ~ & 14.65 & 1.339 & 0.264 \\ 
        21.70 & 1.776 & 0.899 & ~ & 17.46 & 1.246 & 0.730 & ~ & 14.61 & 1.344 & 0.262 \\ 
        21.61 & 1.767 & 0.900 & ~ & 17.40 & 1.239 & 0.721 & ~ & 14.57 & 1.348 & 0.261 \\ 
        21.52 & 1.758 & 0.901 & ~ & 17.34 & 1.232 & 0.711 & ~ & 14.53 & 1.353 & 0.260 \\ 
        21.43 & 1.749 & 0.901 & ~ & 17.29 & 1.226 & 0.701 & ~ & 14.48 & 1.356 & 0.259 \\ 
        21.34 & 1.740 & 0.901 & ~ & 17.23 & 1.220 & 0.691 & ~ & 14.44 & 1.360 & 0.258 \\ 
        21.25 & 1.731 & 0.902 & ~ & 17.17 & 1.214 & 0.681 & ~ & 14.40 & 1.364 & 0.257 \\ 
        21.17 & 1.723 & 0.902 & ~ & 17.11 & 1.208 & 0.670 & ~ & 14.36 & 1.368 & 0.255 \\ 
        21.08 & 1.714 & 0.902 & ~ & 17.06 & 1.203 & 0.660 & ~ & 14.32 & 1.371 & 0.254 \\ 
        20.99 & 1.705 & 0.902 & ~ & 17.00 & 1.198 & 0.649 & ~ & 14.29 & 1.374 & 0.253 \\ 
        20.91 & 1.697 & 0.902 & ~ & 16.95 & 1.194 & 0.638 & ~ & 14.25 & 1.378 & 0.252 \\ 
        20.83 & 1.688 & 0.902 & ~ & 16.89 & 1.189 & 0.627 & ~ & 14.21 & 1.381 & 0.251 \\ 
        20.74 & 1.680 & 0.902 & ~ & 16.84 & 1.186 & 0.616 & ~ & 14.17 & 1.384 & 0.249 \\ 
        20.66 & 1.672 & 0.902 & ~ & 16.78 & 1.182 & 0.604 & ~ & 14.13 & 1.387 & 0.248 \\ 
        20.58 & 1.663 & 0.902 & ~ & 16.73 & 1.179 & 0.593 & ~ & 14.09 & 1.390 & 0.247 \\ 
        20.50 & 1.655 & 0.902 & ~ & 16.67 & 1.177 & 0.581 & ~ & 14.05 & 1.393 & 0.245 \\ 
        20.42 & 1.646 & 0.902 & ~ & 16.62 & 1.174 & 0.570 & ~ & 14.02 & 1.396 & 0.244 \\ 
        20.34 & 1.637 & 0.902 & ~ & 16.57 & 1.172 & 0.558 & ~ & 13.98 & 1.399 & 0.242 \\ 
        20.26 & 1.628 & 0.902 & ~ & 16.51 & 1.171 & 0.547 & ~ & 13.94 & 1.402 & 0.241 \\ 
        20.18 & 1.619 & 0.902 & ~ & 16.46 & 1.170 & 0.535 & ~ & 13.90 & 1.406 & 0.240 \\ 
        20.10 & 1.610 & 0.901 & ~ & 16.41 & 1.169 & 0.524 & ~ & 13.87 & 1.409 & 0.238 \\ 
        20.02 & 1.601 & 0.901 & ~ & 16.36 & 1.169 & 0.513 & ~ & 13.83 & 1.412 & 0.237 \\ 
        19.94 & 1.592 & 0.900 & ~ & 16.31 & 1.169 & 0.501 & ~ & 13.79 & 1.415 & 0.235 \\ 
        19.87 & 1.583 & 0.899 & ~ & 16.26 & 1.170 & 0.490 & ~ & 13.75 & 1.418 & 0.234 \\ 
        19.79 & 1.573 & 0.898 & ~ & 16.20 & 1.171 & 0.479 & ~ & 13.72 & 1.421 & 0.233 \\ 
        19.72 & 1.564 & 0.897 & ~ & 16.15 & 1.172 & 0.468 & ~ & 13.68 & 1.424 & 0.231 \\ 
        19.64 & 1.555 & 0.896 & ~ & 16.10 & 1.174 & 0.457 & ~ & 13.65 & 1.428 & 0.230 \\ 
        13.61 & 1.431 & 0.229 & ~ & 11.73 & 1.791 & 0.327 & ~ & 10.31 & 1.534 & 0.874 \\ 
        13.57 & 1.434 & 0.227 & ~ & 11.71 & 1.795 & 0.339 & ~ & 10.29 & 1.524 & 0.879 \\ 
        13.54 & 1.437 & 0.226 & ~ & 11.68 & 1.800 & 0.350 & ~ & 10.27 & 1.514 & 0.884 \\ 
        13.50 & 1.440 & 0.225 & ~ & 11.65 & 1.803 & 0.362 & ~ & 10.25 & 1.503 & 0.890 \\ 
        13.47 & 1.443 & 0.223 & ~ & 11.63 & 1.806 & 0.374 & ~ & 10.23 & 1.493 & 0.895 \\ 
        13.43 & 1.447 & 0.222 & ~ & 11.60 & 1.808 & 0.386 & ~ & 10.21 & 1.482 & 0.900 \\ 
        13.40 & 1.450 & 0.221 & ~ & 11.58 & 1.810 & 0.397 & ~ & 10.19 & 1.471 & 0.905 \\ 
        13.36 & 1.453 & 0.219 & ~ & 11.55 & 1.811 & 0.409 & ~ & 10.17 & 1.460 & 0.910 \\ 
        13.33 & 1.457 & 0.218 & ~ & 11.52 & 1.812 & 0.421 & ~ & 10.15 & 1.449 & 0.915 \\ 
        13.30 & 1.460 & 0.216 & ~ & 11.50 & 1.812 & 0.433 & ~ & 10.13 & 1.437 & 0.919 \\ 
        13.26 & 1.463 & 0.215 & ~ & 11.47 & 1.812 & 0.444 & ~ & 10.11 & 1.425 & 0.923 \\ 
        13.23 & 1.467 & 0.213 & ~ & 11.45 & 1.811 & 0.455 & ~ & 10.09 & 1.413 & 0.927 \\ 
        13.19 & 1.471 & 0.211 & ~ & 11.42 & 1.810 & 0.466 & ~ & 10.07 & 1.400 & 0.931 \\ 
        13.16 & 1.474 & 0.210 & ~ & 11.40 & 1.809 & 0.477 & ~ & 10.05 & 1.388 & 0.934 \\ 
        13.13 & 1.478 & 0.208 & ~ & 11.37 & 1.807 & 0.488 & ~ & 10.03 & 1.375 & 0.937 \\ 
        13.09 & 1.482 & 0.207 & ~ & 11.35 & 1.805 & 0.498 & ~ & 10.01 & 1.362 & 0.940 \\ 
        13.06 & 1.486 & 0.205 & ~ & 11.32 & 1.802 & 0.508 & ~ & 9.99 & 1.349 & 0.942 \\ 
        13.03 & 1.491 & 0.204 & ~ & 11.30 & 1.799 & 0.518 & ~ & 9.97 & 1.335 & 0.944 \\ 
        13.00 & 1.495 & 0.202 & ~ & 11.27 & 1.797 & 0.528 & ~ & 9.95 & 1.322 & 0.945 \\ 
        12.96 & 1.500 & 0.201 & ~ & 11.25 & 1.794 & 0.537 & ~ & 9.93 & 1.308 & 0.947 \\ 
        12.93 & 1.504 & 0.200 & ~ & 11.22 & 1.791 & 0.546 & ~ & 9.91 & 1.295 & 0.948 \\ 
        12.90 & 1.509 & 0.198 & ~ & 11.20 & 1.788 & 0.555 & ~ & 9.90 & 1.281 & 0.948 \\ 
        12.87 & 1.514 & 0.197 & ~ & 11.18 & 1.785 & 0.564 & ~ & 9.88 & 1.267 & 0.948 \\ 
        12.84 & 1.519 & 0.196 & ~ & 11.15 & 1.782 & 0.573 & ~ & 9.86 & 1.253 & 0.948 \\ 
        12.80 & 1.524 & 0.195 & ~ & 11.13 & 1.780 & 0.581 & ~ & 9.84 & 1.239 & 0.947 \\ 
        12.77 & 1.530 & 0.193 & ~ & 11.10 & 1.777 & 0.590 & ~ & 9.82 & 1.225 & 0.946 \\ 
        12.74 & 1.535 & 0.193 & ~ & 11.08 & 1.774 & 0.599 & ~ & 9.80 & 1.211 & 0.944 \\ 
        12.71 & 1.541 & 0.192 & ~ & 11.06 & 1.771 & 0.609 & ~ & 9.78 & 1.197 & 0.942 \\ 
        12.68 & 1.547 & 0.191 & ~ & 11.03 & 1.768 & 0.618 & ~ & 9.77 & 1.183 & 0.940 \\ 
        12.65 & 1.553 & 0.190 & ~ & 11.01 & 1.764 & 0.627 & ~ & 9.75 & 1.169 & 0.937 \\ 
        12.62 & 1.559 & 0.190 & ~ & 10.99 & 1.760 & 0.637 & ~ & 9.73 & 1.155 & 0.934 \\ 
        12.59 & 1.565 & 0.190 & ~ & 10.96 & 1.756 & 0.646 & ~ & 9.71 & 1.141 & 0.931 \\ 
        12.56 & 1.572 & 0.190 & ~ & 10.94 & 1.751 & 0.656 & ~ & 9.69 & 1.128 & 0.927 \\ 
        12.53 & 1.579 & 0.190 & ~ & 10.92 & 1.746 & 0.665 & ~ & 9.67 & 1.114 & 0.922 \\ 
        12.50 & 1.586 & 0.190 & ~ & 10.89 & 1.741 & 0.674 & ~ & 9.66 & 1.100 & 0.918 \\ 
        12.47 & 1.593 & 0.191 & ~ & 10.87 & 1.736 & 0.683 & ~ & 9.64 & 1.087 & 0.912 \\ 
        12.44 & 1.600 & 0.192 & ~ & 10.85 & 1.730 & 0.692 & ~ & 9.62 & 1.073 & 0.907 \\ 
        12.41 & 1.607 & 0.193 & ~ & 10.83 & 1.724 & 0.701 & ~ & 9.60 & 1.060 & 0.901 \\ 
        12.38 & 1.614 & 0.194 & ~ & 10.80 & 1.718 & 0.709 & ~ & 9.58 & 1.047 & 0.894 \\ 
        12.35 & 1.622 & 0.195 & ~ & 10.78 & 1.712 & 0.717 & ~ & 9.57 & 1.034 & 0.887 \\ 
        12.32 & 1.630 & 0.197 & ~ & 10.76 & 1.705 & 0.726 & ~ & 9.55 & 1.022 & 0.880 \\ 
        12.29 & 1.637 & 0.199 & ~ & 10.74 & 1.699 & 0.734 & ~ & 9.53 & 1.010 & 0.872 \\ 
        12.26 & 1.645 & 0.202 & ~ & 10.71 & 1.692 & 0.741 & ~ & 9.51 & 0.998 & 0.864 \\ 
        12.23 & 1.653 & 0.204 & ~ & 10.69 & 1.686 & 0.749 & ~ & 9.50 & 0.986 & 0.855 \\ 
        12.20 & 1.661 & 0.208 & ~ & 10.67 & 1.679 & 0.757 & ~ & 9.48 & 0.975 & 0.847 \\ 
        12.17 & 1.669 & 0.211 & ~ & 10.65 & 1.672 & 0.765 & ~ & 9.46 & 0.965 & 0.838 \\ 
        12.14 & 1.678 & 0.215 & ~ & 10.63 & 1.665 & 0.773 & ~ & 9.45 & 0.955 & 0.828 \\ 
        12.12 & 1.686 & 0.219 & ~ & 10.60 & 1.658 & 0.781 & ~ & 9.43 & 0.945 & 0.819 \\ 
        12.09 & 1.694 & 0.224 & ~ & 10.58 & 1.650 & 0.788 & ~ & 9.41 & 0.936 & 0.810 \\ 
        12.06 & 1.703 & 0.229 & ~ & 10.56 & 1.643 & 0.796 & ~ & 9.39 & 0.926 & 0.801 \\ 
        12.03 & 1.711 & 0.234 & ~ & 10.54 & 1.634 & 0.803 & ~ & 9.38 & 0.917 & 0.791 \\ 
        12.00 & 1.719 & 0.240 & ~ & 10.52 & 1.626 & 0.811 & ~ & 9.36 & 0.909 & 0.782 \\ 
        11.98 & 1.727 & 0.246 & ~ & 10.50 & 1.618 & 0.818 & ~ & 9.34 & 0.900 & 0.772 \\ 
        11.95 & 1.735 & 0.253 & ~ & 10.48 & 1.609 & 0.825 & ~ & 9.33 & 0.892 & 0.762 \\ 
        11.92 & 1.743 & 0.261 & ~ & 10.45 & 1.600 & 0.832 & ~ & 9.31 & 0.884 & 0.752 \\ 
        11.89 & 1.751 & 0.269 & ~ & 10.43 & 1.591 & 0.838 & ~ & 9.29 & 0.876 & 0.742 \\ 
        11.87 & 1.759 & 0.277 & ~ & 10.41 & 1.581 & 0.844 & ~ & 9.28 & 0.869 & 0.732 \\ 
        11.84 & 1.766 & 0.286 & ~ & 10.39 & 1.572 & 0.851 & ~ & 9.26 & 0.862 & 0.721 \\ 
        11.81 & 1.773 & 0.296 & ~ & 10.37 & 1.563 & 0.857 & ~ & 9.24 & 0.855 & 0.711 \\ 
        11.79 & 1.779 & 0.306 & ~ & 10.35 & 1.553 & 0.862 & ~ & 9.23 & 0.849 & 0.700 \\ 
        11.76 & 1.785 & 0.316 & ~ & 10.33 & 1.543 & 0.868 & ~ & 9.21 & 0.843 & 0.690 \\ 
        9.19 & 0.837 & 0.679 & ~ & 8.74 & 0.777 & 0.372 & ~ & 8.34 & 0.870 & 0.150 \\ 
        9.18 & 0.831 & 0.668 & ~ & 8.73 & 0.779 & 0.363 & ~ & 8.32 & 0.875 & 0.145 \\ 
        9.16 & 0.826 & 0.658 & ~ & 8.71 & 0.781 & 0.355 & ~ & 8.31 & 0.881 & 0.140 \\ 
        9.15 & 0.821 & 0.647 & ~ & 8.70 & 0.783 & 0.346 & ~ & 8.30 & 0.886 & 0.136 \\ 
        9.13 & 0.816 & 0.637 & ~ & 8.69 & 0.784 & 0.338 & ~ & 8.28 & 0.891 & 0.132 \\ 
        9.11 & 0.812 & 0.626 & ~ & 8.67 & 0.786 & 0.329 & ~ & 8.27 & 0.896 & 0.128 \\ 
        9.10 & 0.807 & 0.615 & ~ & 8.66 & 0.788 & 0.321 & ~ & 8.26 & 0.901 & 0.124 \\ 
        9.08 & 0.803 & 0.605 & ~ & 8.64 & 0.790 & 0.313 & ~ & 8.24 & 0.906 & 0.120 \\ 
        9.07 & 0.799 & 0.594 & ~ & 8.63 & 0.791 & 0.305 & ~ & 8.23 & 0.911 & 0.117 \\ 
        9.05 & 0.795 & 0.583 & ~ & 8.61 & 0.793 & 0.297 & ~ & 8.22 & 0.916 & 0.113 \\ 
        9.03 & 0.792 & 0.572 & ~ & 8.60 & 0.795 & 0.289 & ~ & 8.20 & 0.921 & 0.110 \\ 
        9.02 & 0.789 & 0.562 & ~ & 8.59 & 0.797 & 0.280 & ~ & 8.19 & 0.925 & 0.107 \\ 
        9.00 & 0.786 & 0.551 & ~ & 8.57 & 0.799 & 0.272 & ~ & 8.18 & 0.930 & 0.104 \\ 
        8.99 & 0.783 & 0.540 & ~ & 8.56 & 0.801 & 0.264 & ~ & 8.17 & 0.935 & 0.101 \\ 
        8.97 & 0.780 & 0.529 & ~ & 8.54 & 0.804 & 0.255 & ~ & 8.15 & 0.939 & 0.099 \\ 
        8.96 & 0.778 & 0.518 & ~ & 8.53 & 0.807 & 0.247 & ~ & 8.14 & 0.944 & 0.096 \\ 
        8.94 & 0.776 & 0.507 & ~ & 8.51 & 0.810 & 0.238 & ~ & 8.13 & 0.948 & 0.094 \\ 
        8.92 & 0.775 & 0.496 & ~ & 8.50 & 0.813 & 0.230 & ~ & 8.11 & 0.952 & 0.092 \\ 
        8.91 & 0.773 & 0.485 & ~ & 8.49 & 0.817 & 0.222 & ~ & 8.10 & 0.956 & 0.089 \\ 
        8.89 & 0.772 & 0.475 & ~ & 8.47 & 0.821 & 0.214 & ~ & 8.09 & 0.960 & 0.087 \\ 
        8.88 & 0.771 & 0.464 & ~ & 8.46 & 0.825 & 0.206 & ~ & 8.08 & 0.965 & 0.085 \\ 
        8.86 & 0.771 & 0.453 & ~ & 8.45 & 0.830 & 0.199 & ~ & 8.06 & 0.968 & 0.083 \\ 
        8.85 & 0.771 & 0.442 & ~ & 8.43 & 0.834 & 0.192 & ~ & 8.05 & 0.972 & 0.081 \\ 
        8.83 & 0.771 & 0.432 & ~ & 8.42 & 0.839 & 0.185 & ~ & 8.04 & 0.976 & 0.080 \\ 
        8.82 & 0.771 & 0.421 & ~ & 8.40 & 0.844 & 0.178 & ~ & 8.03 & 0.980 & 0.078 \\ 
        8.80 & 0.772 & 0.411 & ~ & 8.39 & 0.849 & 0.172 & ~ & 8.01 & 0.984 & 0.076 \\ 
        8.79 & 0.773 & 0.401 & ~ & 8.38 & 0.854 & 0.166 & ~ & 8.00 & 0.987 & 0.075 \\ 
        8.77 & 0.774 & 0.391 & ~ & 8.36 & 0.860 & 0.160 & ~ & ~ & ~ & ~ \\ 
        8.76 & 0.776 & 0.382 & ~ & 8.35 & 0.865 & 0.155 & ~ & ~ & ~ & ~ \\ 
\end{longtable}
\end{appendix}
\end{document}